\documentclass[sn-mathphys]{sn-jnl}

\usepackage{epsfig}
\usepackage{bm}
\usepackage{amsmath,amssymb,amsfonts}
\usepackage{xcolor}
\usepackage{manyfoot}
\usepackage{graphicx}
\usepackage{multirow}
\usepackage{amsthm}
\usepackage{textcomp}

\begin{document}

\title{Galactic foreground and CMB emissions randomization due to chaotic/turbulent  dynamics of magnetized plasma dominated by magnetic helicity}

\author*{\fnm{Alexander} \sur{Bershadskii}}
\email{bershads@gmail.com}

\affil{\orgname{ICAR}, \orgaddress{\city{P.O. Box 31155, Jerusalem}, \postcode{91000}, \country{Israel}}}

\abstract{
  Using results of numerical simulations and astrophysical observations (mainly in the WMAP and Planck frequency bands) it is shown that Galactic foreground emission becomes more sensitive to the mean magnetic field with the frequency, that results in the appearance of two levels of its randomization due to chaotic/turbulent dynamics of magnetized interstellar medium dominated by the magnetic helicity. The galactic foreground emission is more randomized at higher frequencies. The Galactic synchrotron and polarized dust emissions have been studied in detail. It is shown that the magnetic field imposes its level of randomization on the synchrotron and dust emission. The background magnetic field (around the lepto/baryogenesis) and CMB emission have also been briefly discussed in this context. It is shown that they are considerably less randomized than the foreground ones. The main method for the theoretical consideration used in this study is the Kolmogorov-Iroshnikov phenomenology within the framework of the distributed chaos notion.  Despite the vast differences in the values of physical parameters and spatio-temporal scales between the numerical simulations and the astrophysical observations, there is a quantitative agreement between the results of the astrophysical observations and the numerical simulations within the framework of the distributed chaos notion.}

\maketitle

\newpage

\section{Introduction}

  The role of Galactic foreground emission in astrophysics is twofold. On the one hand, different types of Galactic emission contributing to the Galactic foreground are indispensable sources of information about physical processes in the magnetized interstellar medium. On the other hand, the Galactic foreground is the main obstacle to obtaining a clean cosmic microwave background (CMB) radiation map which is the main observational source of information about physical processes at the early stages of the universe's development. Therefore, investigation of the Galactic foreground emission (and its components) is necessary to solve the two important problems of modern astrophysics. 
  
  The recent major satellite missions: WMAP and Planck, were designed mainly to solve the second problem. However, to solve this problem we need an effective method for separating the CMB and the foreground (mainly of Galactic origin). Such techniques were developed in the last decades. The maps of the most important components of the Galactic foreground (such as synchrotron and dust emission) were also obtained as a necessary by-product of this activity. \\

    The main difficulty in interpreting and understanding these results is a weakness of the theory of the chaotic/turbulent processes in the interstellar magnetized medium. These processes are supposed to be the main physical source of the apparently random character of the foreground maps. The scaling (power-law) approach, widely used for interpreting the power spectra corresponding to the maps, requires a wide range of scales for its validation which is rarely achievable in practice.   \\
    
    The conception of smoothness can be used instead to quantify the levels of randomness of the chaotic/turbulent dynamical regimes. Indeed, the stretched exponential spectrum
\begin{equation}
E(k) \propto \exp-(k/k_{\beta})^{\beta} 
\end{equation}
is a characteristic feature of smooth chaotic dynamics. Here $1 \geq \beta > 0$ and $k$ is the wavenumber. 

The value of the $\beta =1$ characterizes the deterministic chaos (see, for instance, \citealt{mm1, mm2, mm3, kds} and references therein):
\begin{equation}
E(k) \propto \exp(-k/k_c).  
\end{equation}

When $1 > \beta$ the smooth chaotic dynamics can be already non-deterministic, this type of smooth dynamics can be called `distributed chaos' (the term will be clarified below). Another term ``soft turbulence'' (suggested by \citealt{wu}) can be also appropriate.\\

 The parameter $\beta$ could be used as an informative measure of randomization. Namely, the further the value of  $\beta$ is from the $\beta =1$ (which corresponds to the deterministic chaos) the more significant the system's randomization. The smaller parameter $\beta$ values are considered a precursor of hard turbulence. The scaling power spectrum is a characteristic feature of non-smooth random dynamics (the hard turbulence in terms of \citealt{wu}).\\

  Figures 1 and 2 (adapted from Figs. B1 and B5 of a paper \citealp{gho}) show the full-sky Galactic foreground maps computed using the data measured by the probes onboard the WMAP satellite for  K and W frequency bands. \\

\begin{figure} \vspace{-2.2cm}\centering \hspace{-1.1cm}
\epsfig{width=.7\textwidth,file=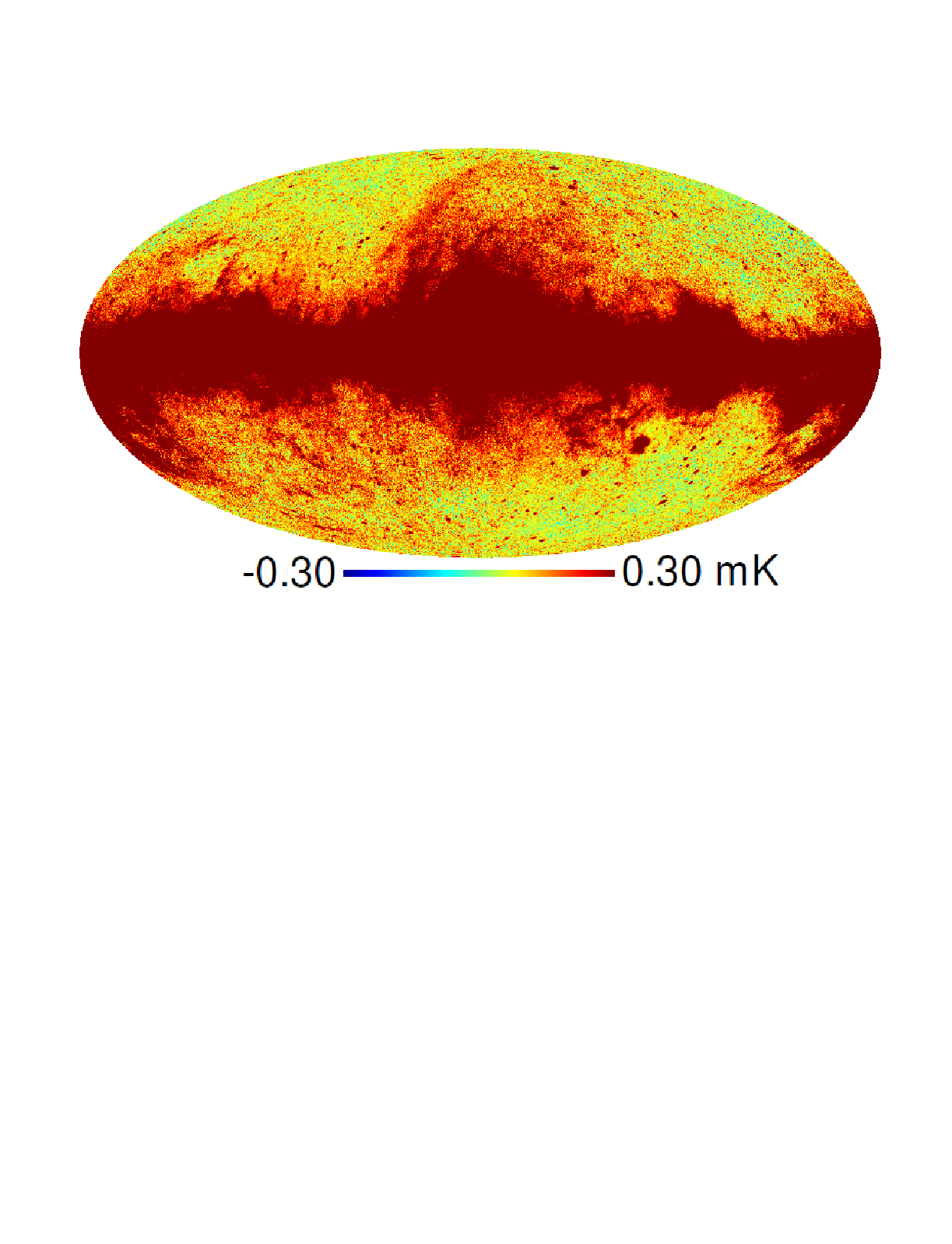} \vspace{-5.9cm}
\caption{Full-sky Galactic foreground map for the WMAP K-band (central frequency 23 GHz).} 
\end{figure}
\begin{figure} \vspace{-0.7cm}\centering \hspace{-1.1cm}
\epsfig{width=.7\textwidth,file=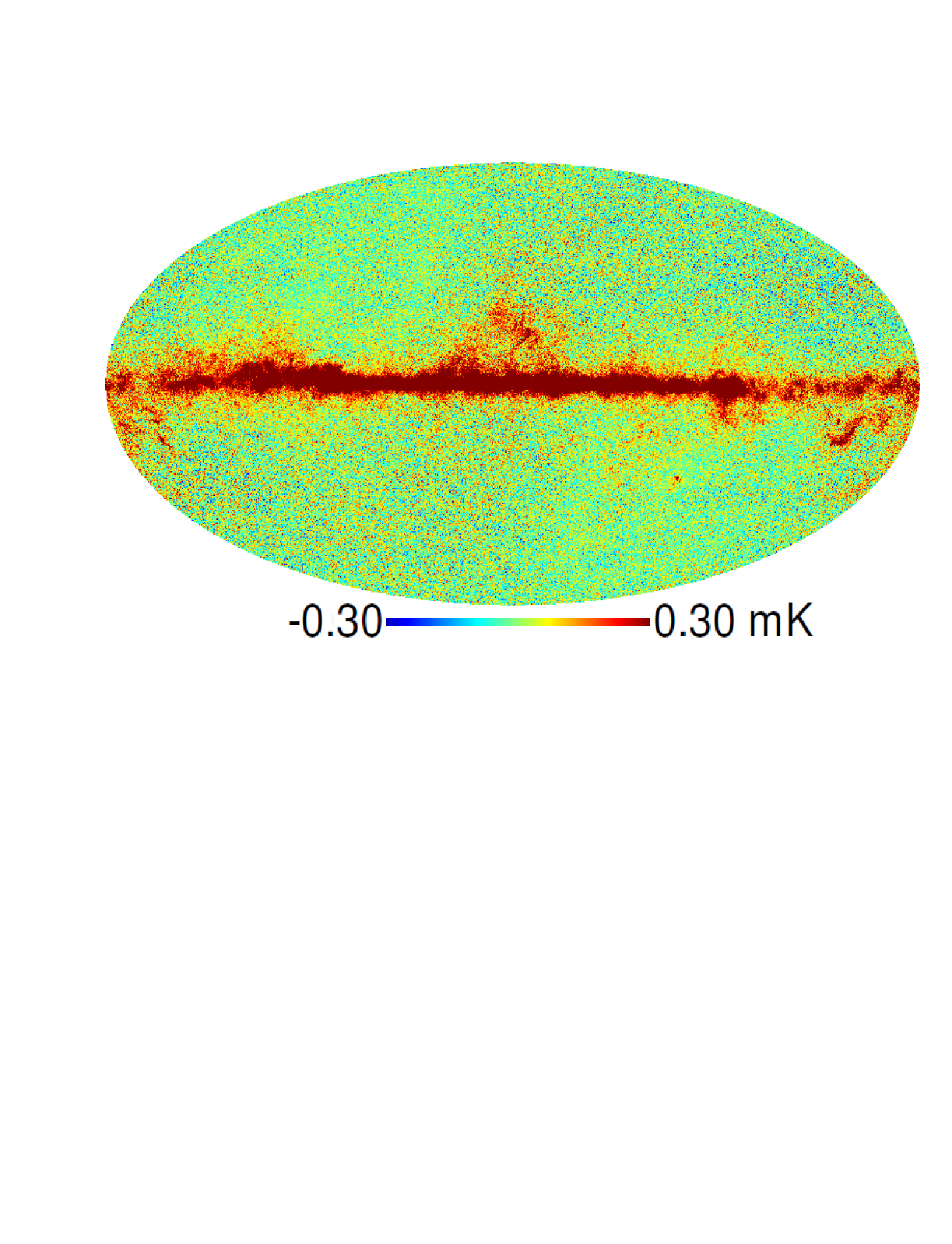} \vspace{-5.4cm}
\caption{Full-sky Galactic foreground map for the WMAP W-band ((central frequency 94 GHz).}
\end{figure}
\begin{figure} \vspace{-2.3cm}\centering 
\epsfig{width=.7\textwidth,file=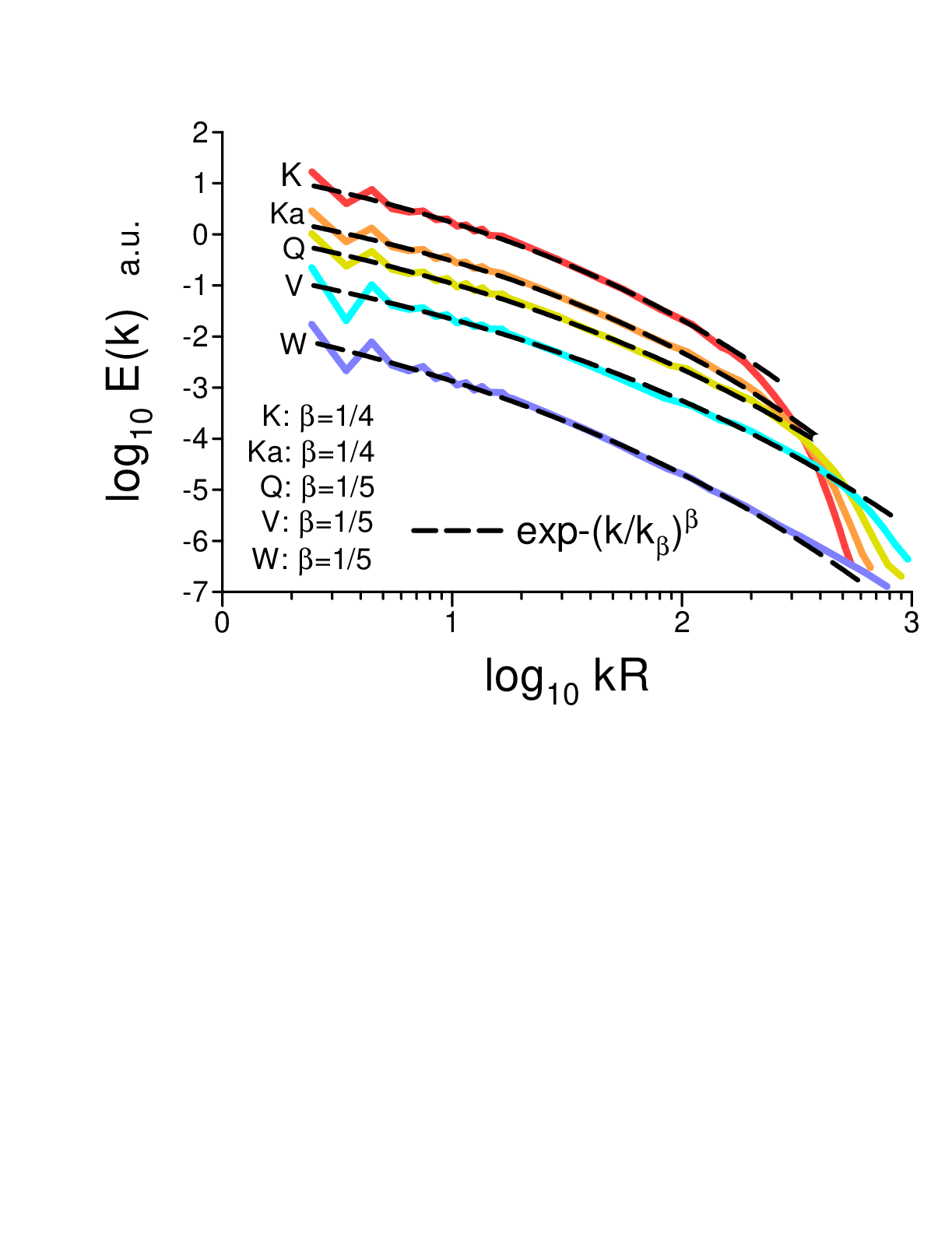} \vspace{-5.1cm}
\caption{Power spectra corresponding to the full-sky Galactic foreground maps for all WMAP frequency bands.} 
\end{figure}

  Figure 3 shows the power spectra corresponding to the full-sky Galactic foreground maps for the K - central frequency 23 GHz, Ka - central frequency 33 GHz, Q - central frequency 41 GHz, V - central frequency 61 GHz, W - central frequency 94 GHz frequency bands. The spectral data for Fig. 3 were taken from Fig. 8 of the paper \citep{gho}. In the original figure in the paper \citep{gho}, the angular power spectra $C_l$ are shown vs the multipole $l$. The spherical multipole $l$ can be related to a wavenumber $k_l = \sqrt{l(l+1)}/R$, where $R$ is the sphere's radius, and the azimuthally averaged 2-D power spectral density $E(k_l) \approx R^2 C_l/\pi$ \citep{maus}.\\
   
  The dashed curves in Fig. 3 indicate the best fit by the stretched exponential spectrum Eq. (1) (the distributed chaos). One can see that $\beta=1/4$ for the frequency bands K and Ka whereas $\beta =1/5$ for the frequency bands Q, V, and W. These values of $\beta$ will be explained below. Now we would like to emphasize that there is a two-level randomization in the WMAP  foreground maps (the two values of $\beta$) and the higher level of randomization (smaller $\beta$) corresponds to higher frequencies.\\
  
    The present paper will relate the apparent two-level randomization of the Galactic foreground emission to the Kolmogorov-Iroshnikov phenomenology \citep{my,ir} applied to the magnetic helicity-dominated chaotic/turbulent motion of the magnetized interstellar medium in the frames of distributed chaos  (the magneto-inertial range of scales \citealp{ber4}). A comparison with the randomization of the cosmic microwave background (CMB) will be briefly discussed and it will be shown that the CMB emission is considerably less randomized ($\beta = 1/2$) than the Galactic foreground emission ($\beta = 1/4,~1/5$). 
   
\section{Magnetic helicity and distributed chaos}  

\subsection{Deterministic chaos in magnetized plasma}

The estimates of the values of the Galactic magnetic field obtained by the observations turned out to be considerably larger than the values predicted for the primordial magnetic field. Therefore certain mechanisms of amplification of the magnetic fields by the intense chaotic/turbulent motion of the electrically conducting Galactic plasma (dynamo) were suggested in the existing literature (see \citealp{rss,sok,vaz,sub,kor} and references therein). This motion could be produced, by the Galactic differential rotation and the supernova explosions. 
 
  In a recent paper \citep{seta1} a numerical simulation of a small-scale (fluctuating) dynamo with parameters favorable to deterministic chaos was performed using magnetohydrodynamic (MHD) equations
 
\begin{align}
\frac{\partial \rho}{\partial t} & + \nabla \cdot (\rho {\bf u})  = 0,  \\
\frac{\partial {\bf b}}{\partial t} & = \nabla \times ({\bf u} \times {\bf b}) + \eta \nabla^2 {\bf b},  \\ 
\frac{\partial {\bf u}}{\partial t} & + ({\bf u} \cdot \nabla) {\bf u} = -\frac{\nabla p}{\rho} + \frac{ {\bf j} \times {\bf b}}{c\rho}  \nonumber \\
& + \nu \left(\nabla^2 {\bf u} + \frac{1}{3} \nabla (\nabla \cdot {\bf u}) + 2 {\bf S} \cdot \nabla \ln \rho \right) + {\bf F}, 
\end{align}
in a triply-periodic cubic domain. In these equations ${\bf u}$ is the plasma velocity field, ${\bf b}$ is the divergence-free magnetic field, $\rho$ is the plasma density, $p$ is the plasma pressure, $\nu$ is the plasma viscosity, $\eta$ is the plasma magnetic diffusivity, ${\bf j} = (c/4\pi)\nabla\times {\bf b}$ was taken for electric current density, $c$ is the speed of light, $S_{ij} = \frac{1}{2} \left(u_{i,j} + u_{j,i} - \frac{2}{3} \delta_{ij} \nabla \cdot \bf{u} \right)$ was taken for the rate-of-strain tensor, and ${\bf F}$ is a random delta-correlated in time solenoidal forcing function. An isothermal equation of state, $ p =c_s^2 \rho $, was assumed with constant sound speed $c_s$. 

\begin{figure} \vspace{-2.1cm}\centering 
\epsfig{width=.65\textwidth,file=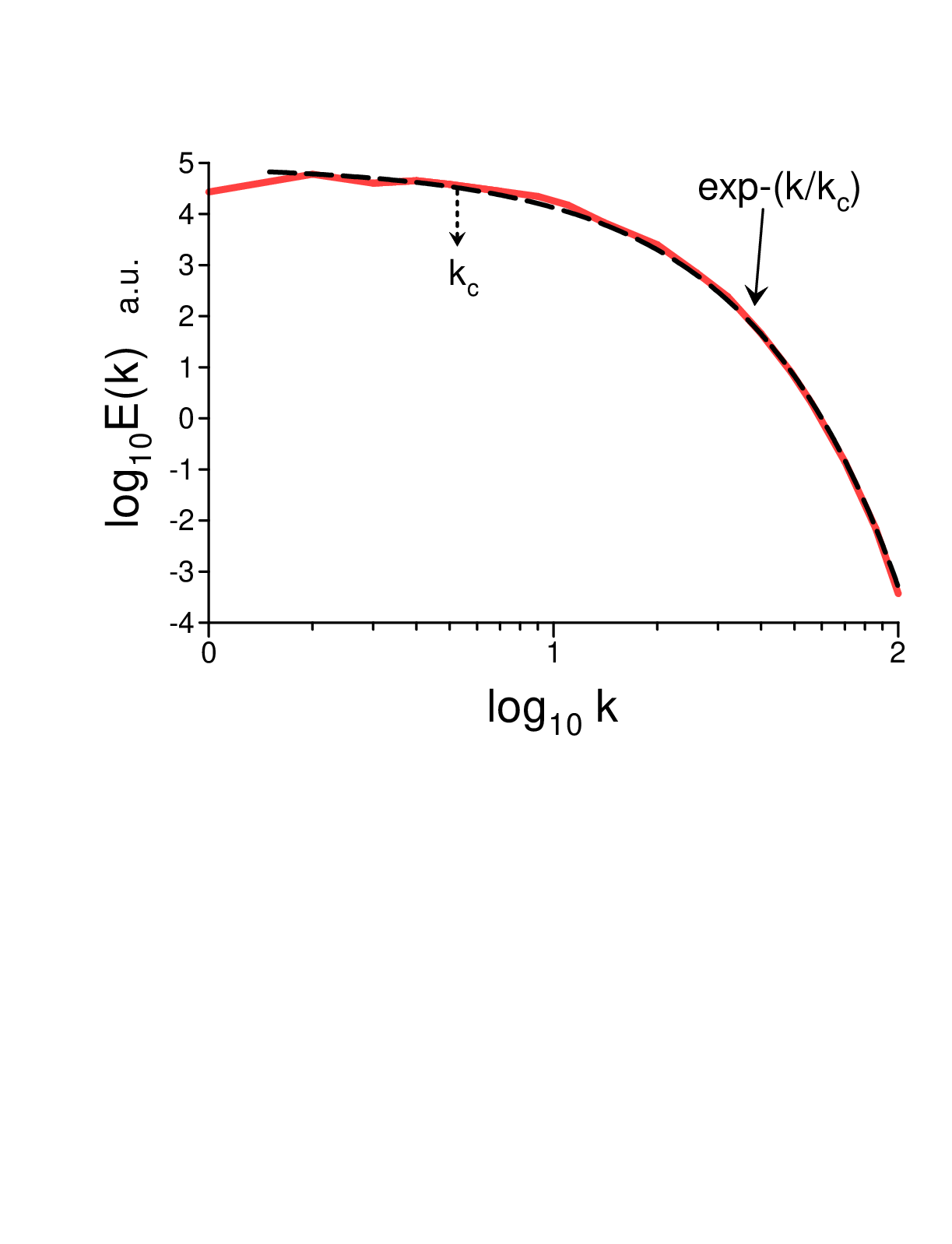} \vspace{-4.4cm}
\caption{Magnetic energy spectrum at the saturated stage of the MHD dynamo for the Mach number $M  \approx 0.11$.} 
\end{figure}

  The Reynolds and magnetic Reynolds numbers $Re =Re_m =1122$, the Mach number $M  \approx 0.11$, and the magnetic Prandtl number $Pr_m =1$. A weak random magnetic field (with zero net flux across the computational domain) was used as an initial seed field.\\
  
    Fig. 4 shows the one-dimensional (shell-averaged)  magnetic energy spectrum at the saturated stage of the dynamo (the spectral data were taken from Fig. 2 of the paper \citealp{seta1}). The dashed curve is the best fit corresponding to Eq. (2) (deterministic chaos).
  
 \subsection{Magnetic helicity}   
 
 The ideal MHD has three fundamental quadratic invariants: total energy, cross and magnetic helicity \citep{mt}.  The validity of magnetic helicity conservation increases with the magnetic Reynolds number value $Re_m$. The Galaxy magnetized plasma is characterized by very large magnetic Reynolds numbers  \citep{zv}.\\
  
  The average magnetic helicity density is
\begin{equation}
 h_m = \langle {\bf a} {\bf b} \rangle   
\end{equation}
here ${\bf a}$ is the vector potential, ${\bf b} = [{\nabla \times \bf a}]$ is the fluctuating magnetic field, and $\langle ... \rangle$ means spatial average (for the fluctuating variables  $\langle {\bf a} \rangle = \langle {\bf b} \rangle = 0$). \\

  The magnetic helicity is not invariant in the uniform mean magnetic field ${\bf B_0}$. However, a generalized average magnetic helicity density
 \begin{equation}
 \hat{h}_m = h_m + 2{\bf B_0}\cdot \langle {\bf A}  \rangle 
\end{equation}
where ${\bf B} = {\bf B_0} + {\bf b}$, ${\bf A} = {\bf A_0} +{\bf a}$, is still an ideal invariant \citep{mg}
\begin{equation}
 \frac{d \hat{h}_m}{d t} =  0
\end{equation}
(see also \citealp{shebalin}).\\
  
   The magnetic helicity can be considered as an adiabatic invariant not only in the ideal MHD but also in a weakly dissipative magnetized plasma (see for instance, \citealt{zv}), that makes it especially interesting for the interstellar media.
   
\subsection{Distributed chaos dominated by magnetic helicity} 
   
   The transition from deterministic chaos to distributed one can be considered as a randomization. Namely,  the change of physical parameters can result in the random fluctuations of the characteristic scale $k_c$ in equation (2). One has to take this phenomenon into account. It can be done using an ensemble averaging 
\begin{equation}
E(k) \propto \int_0^{\infty} P(k_c) \exp -(k/k_c)dk_c 
\end{equation}  

   Here, a probability {\it distribution} $P(k_c)$ describes the random fluctuations of $k_c$. This is the rationale behind the name `distributed chaos'.

  For the magnetic field dynamics dominated by the magnetic helicity the scaling relationship between characteristic values of $k_c$ and $B_c$ based on dimensional considerations
 \begin{equation}
B_c \propto |h_m|^{1/2} k_c^{1/2}   
\end{equation}
can be used to find the probability distribution $P(k_c)$.\\

  The value of $B_c$ can be taken half-normally distributed $P(B_c) \propto \exp- (B_c^2/2\sigma^2)$ (\citealt{my}). It is a normal distribution with zero mean which is truncated to have a nonzero probability density function for positive values of its argument only. For instance, if $B$ is a normally distributed random variable, then the variable $B_c = |B|$ is half-normally distributed (\citealt{jkb}). \\
  
    From Eq. (10) we then obtain 
\begin{equation}
P(k_c) \propto k_c^{-1/2} \exp-(k_c/4k_{\beta})  
\end{equation}
 It is the chi-squared probability distribution where $k_{\beta}$ is a new constant). \\

   Substituting Eq. (11) into Eq. (9) one obtains
\begin{equation}
E(k) \propto \exp-(k/k_{\beta})^{1/2}  
\end{equation}

When a non-zero mean magnetic field ${\bf B}_0$ is significant the $h_m$ should be replaced by the generalized averaged magnetic helicity density $\hat{h}_m$ Eq. (7) in the estimate Eq. (10).  The $\hat{h}_m$ has the same dimensionality as $h_m$ and therefore, the magnetic energy spectrum will be given by the same Eq. (12).
  
\section{Parity violation and background radiation} 

\subsection{Magnetic measure of parity violation} 

   To understand the differences between the foreground and background emission randomization, let us briefly describe the background emission origin and randomization. \\
   
    Baryogenesis at an earlier stage of the universe's development is one of the main candidates for magnetogenesis. The parity P (reflection symmetry) and charge-parity (CP) violation are a necessary part of the baryogenesis (Sakharov's conditions). Helical magnetic field emerges under these conditions. It should be noted that the electro-weak phase transition can also be involved in generating a helical magnetic field under these conditions (see, for instance, a review \citealp{sub1} and references therein). 

  The time around the lepto/baryogenesis epoch was rather turbulent \citep{kam} because about all created baryons and antibaryons (as well as electrons and positrons) were annihilated soon after creation, producing a huge amount of energy.  The turbulent magnetic helicity can also be a key for solving the baryon asymmetry problem \citep{kam}. \\
 
   The point-wise (more precisely, the patch-like) helicity production is an inherent property of the 3D chaotic/turbulent motion of the plasmas/fluids. In such flows, the helicity appears spontaneously (even in the cases of zero net helicity) in pairs of patches having opposite signs of the helicity so that the net helicity remains zero. This phenomenon takes place in both non-magnetic (kinetic helicity) and magnetic (magnetic helicity) cases  \citep{kerr,hk,moff2,sche}. In the latter case, a part of the patches is transformed into long-lived magnetic blobs with sign-definite helicity and reduced dissipation \cite{moff2,ber4}. Despite the net helicity is remaining zero the absolute value of the total normalized sum of the sign-definite helicities of the magnetic blobs $|I^{\pm}|$ ($I^{+} =- I^{-}$) can be used as a measure of spontaneous parity violation.
   
  Now we will study how this measure of parity violation is related to a measure of randomness of the chaotic/turbulent magnetic field (some other types of cosmological parity violation built on intrinsic reflection asymmetry of the particles, one can find in a recent review \citealp{ss} and references therein).\\  

For chaotic/turbulent flows with net reflectional symmetry the net magnetic helicity is equal to zero, whereas the point-wise magnetic helicity is not (because of the spontaneous breaking of the local reflectional symmetry \citealp{kerr,hk}). The spontaneous local symmetry breaking in such flows is accompanied by the emergence of the blobs with non-zero helicity \citep{mt,moff1,moff2,lt}. The magnetic surfaces of these blobs can be defined by the boundary conditions: ${\bf b_n}\cdot {\bf n}=0$, where ${\bf n}$ is a unit normal to the boundary of the blob. 
   
    The sign-defined magnetic helicity of the j-blob can be defined as
\begin{equation}
 H_j^{\pm} = \int_{V_j} ({\bf a} ({\bf x},t) \cdot  {\bf b} ({\bf x},t)) ~ d{\bf x} 
\end{equation}  
  where  (`+'  or `-') denotes the blob's helicity sign. The $H_j^{\pm}$ is an adiabatic invariant \citep{mt} (see also above)
  
  Then we can consider the total sign-defined adiabatic invariant 
\begin{equation}
{\rm I^{\pm}} = \lim_{V \rightarrow  \infty} \frac{1}{V} \sum_j H_{j}^{\pm}   
\end{equation}
 The summation takes into account the blobs with a certain sign only (`+' or `-'), and $V$ is the total volume of the blobs taken into account.  \\

  The adiabatic invariant ${\rm I^{\pm}}$ defined by Eq. (14) can be used as a measure of the parity violation, and can be used instead of the averaged magnetic helicity density $h_m$ in the above estimate Eq. (10) for the special case of the local reflectional symmetry breaking
\begin{equation}
B_c \propto |{\rm I^{\pm}}|^{1/2} k_c^{1/2}   
\end{equation}
The spectrum Eq. (12) can also be obtained for this case.  \\

    In recent papers \citep{sf,seta} numerical simulations similar to that considered in Section 2.1 were performed (the net magnetic helicity was also negligible), but in these simulations the large Mach number $M=10$ was achieved. Figure 5 shows the magnetic energy spectra computed in these numerical simulations. The spectral data shown in this figure were taken from Fig. C4 of the Ref. \citep{seta}.  While for the small Mach number $M=0.1$ the bottom dashed curve in Fig. 5 indicates the exponential spectrum Eq. (2) (deterministic chaos, cf Fig. 1), for the large Mach number $M=10$ the top dashed curve indicates a stretched exponential spectrum Eq. (12) with $\beta = 1/2$.  This indicates the distributed chaos dominated by magnetic helicity, i.e., the spontaneous breaking of local reflectional symmetry.  \\

\begin{figure} \vspace{-2cm}\centering 
\epsfig{width=.68\textwidth,file=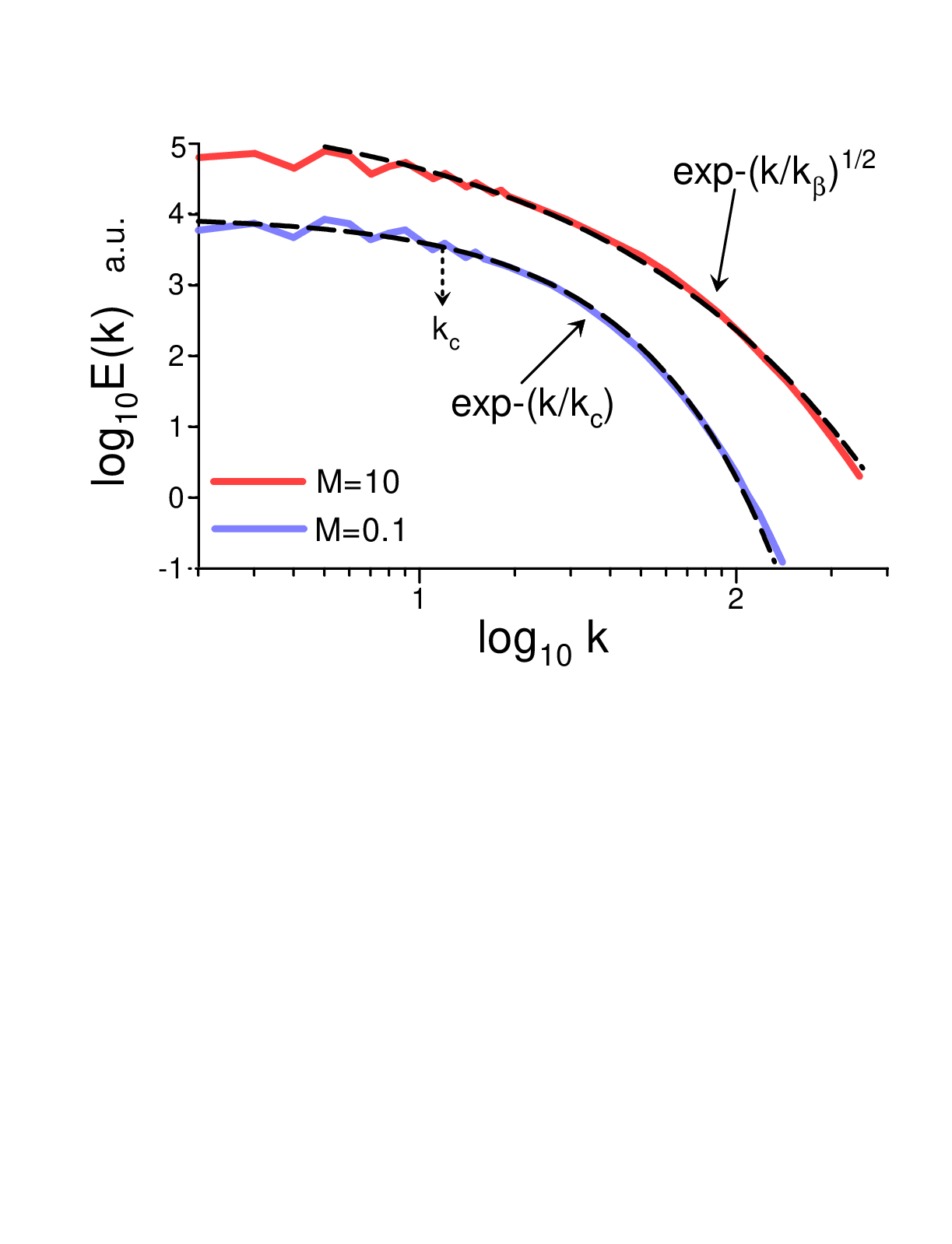} \vspace{-5.3cm}
\caption{Magnetic energy spectra at the saturated stage of the MHD dynamo for the Mach numbers $M =0.1$ (bottom) and $M=10$ (top).}
\end{figure}

\subsection{Magnetic field and background radiation}
   
   Many models of baryogenesis were suggested, but there is a consensus to use the CMB data to estimate the present-day ratio of the number density of baryons ($n_b$) to photons ($n_\gamma$): $\eta_b = n_b/n_\gamma = 6.12\times 10^{-10}$ \citep{agh}. This is the observational key for the baryogenesis problem (see, for a recent review,  \citealp{bb} and references therein).  The extremely low observed value of the parameter $\eta_b$ means, in particular, that the number of photons emitted by the recombination process itself is negligible in comparison with the number of photons that came from the previous epochs and were set free at the recombination epoch.\\
   
  As was already mentioned above, most of the baryons $B$ and antibaryons ${\bar B}$ were annihilated around the baryogenesis epoch, producing a vast amount of energy in the form of photons: $\gamma +\gamma \rightleftarrows B+{\bar B}$. Since these photons in the post-baryogenesis epoch had insufficient energy to turn into the baryon-antibarion pairs, most of them survived until the last scattering to become the CMB photons and provide valuable information about the primordial matter-antimatter asymmetry (the $\eta_b$, in particular) \citep{ba}.  Since the lepto/baryogenesis, magnetogenesis, the parity violation dominated by the magnetic helicity and the emission of the CMB-to-be photons are strongly interrelated one can expect that the level of the randomness of the magnetic field generated at the lepto/baryogenesis epoch could be imprinted on the level of randomness of the temperature of the CMB-to-be photons. \\
   
   The continuing expansion smoothed out the strong fluctuations characterizing the time around the lepto/baryogenesis epoch. At the recombination epoch, only tiny remnants of these fluctuations were present in the matter's motion and CMB photons. The remnants of the randomness, peculiar to the time around the lepto/baryogenesis epoch, could also be preserved in the spatial power spectra of the CMB temperature and polarization fluctuations (anisotropies) after the last scattering. Of course, this spectrum is a superposition of the remnant of the turbulent lepto/baryogenesis epoch, acoustic oscillations, etc. However, one can hope this remnant can be seen in the observed CMB spectra and provide some information about the physical process at the lepto/baryogenesis epoch (see also an interesting discussion of present attempts to approach the problem in a recent paper \citealp{phi} and references therein). \\

\begin{figure} \vspace{-1.5cm}\centering 
\epsfig{width=.65\textwidth,file=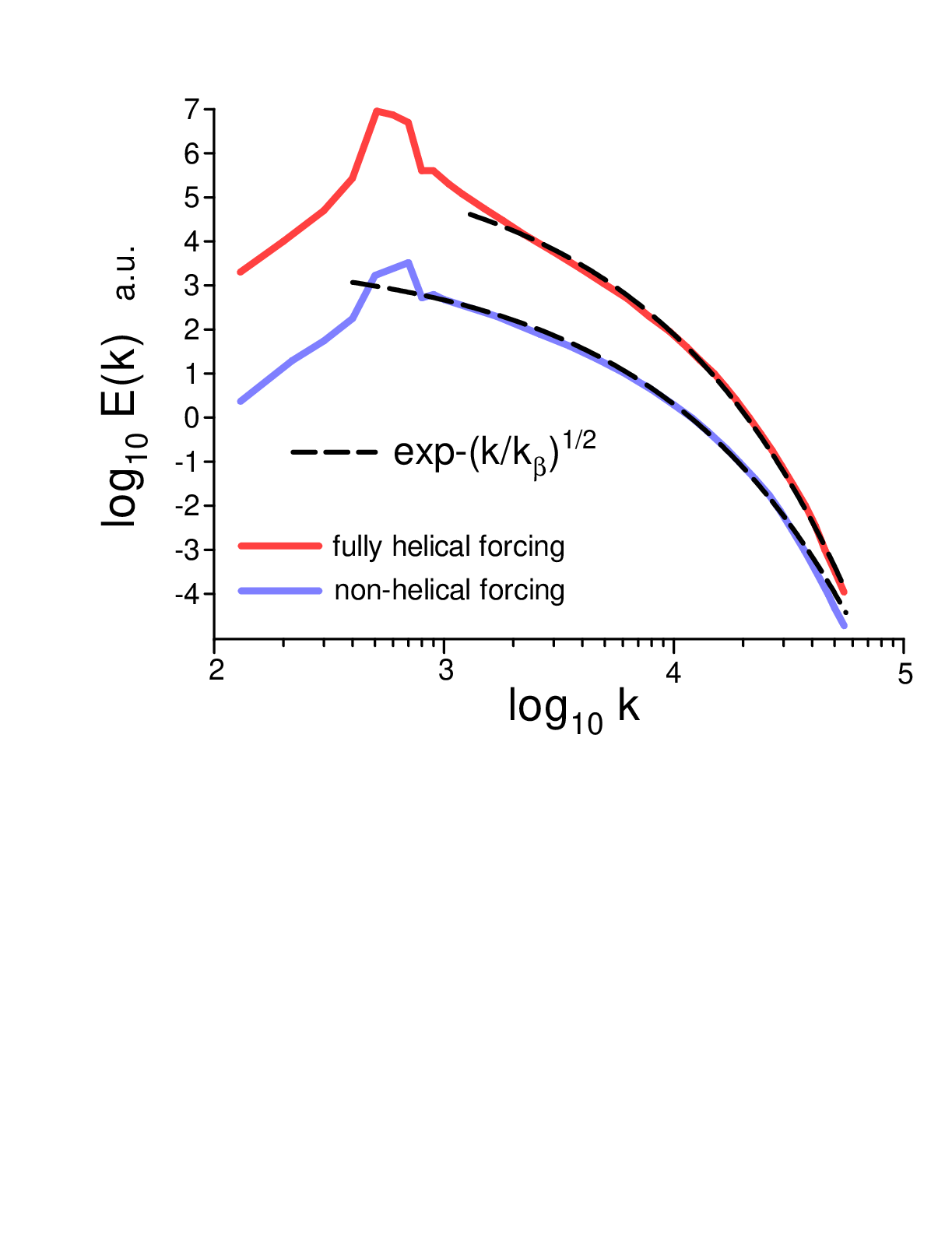} \vspace{-4.4cm}
\caption{Magnetic energy spectra in an expanding flat universe at the radiation-dominated epoch after the electro-weak phase transition (numerical simulations). The spectra are vertically shifted for clarity.}
\end{figure}

   In a recent paper \citep{pol} numerical simulations of magnetized ultrarelativistic plasma in an expanding flat universe at the radiation-dominated epoch after the electro-weak phase transition were performed using the comoving equations
\begin{align}
\!\!{\partial\ln\rho\over\partial t}
\!=\!-\frac{4}{3}\left(\nabla \!\!\cdot \!\!\ {\bf u}\!+\!{\bf u} \! \cdot \!\! \nabla\ln\rho\right)\!\!
+\!\! {1\over\rho} \! \left[{\bf u} \! \cdot \! ({\bf J}\!\times\!{\bf B})\!+\!\eta {\bf J}^2\right]\! , \nonumber\\
{\partial{\bf u}\over\partial t}\!=\!-{\bf u}\cdot\nabla{\bf u}
+{{\bf u}\over3}\left(\nabla\cdot{\bf u}\!+\!{\bf u}\cdot\nabla\ln\rho\right)
+{2\over\rho}\nabla\!\cdot\!\left(\rho\nu{\bf S}\right)  \nonumber\\
-{1\over4}\nabla\ln\rho -{{\bf u}\over\rho}\left[{\bf u}\cdot({\bf J}\times{\bf B})+\eta{\bf J}^2\right] +{3\over4\rho}{\bf J}\times{\bf B}, \nonumber\\
{\partial{\bf B}\over\partial t}=\nabla\times({\bf u}\times{\bf B}-\eta{\bf J}+{\bf F}). \nonumber
\end{align}

The random electromotive force ${\bf F}$ simulates the generation of magnetic fields and was activated for a short duration (10\% of the Hubble time). Two forcing cases were considered: fully helical and with negligible helicity. The simulations were performed in a periodic cubic box. The magnetic Prandtl number $Pr_m = \nu/\eta =1$. \\

  The magnetic energy spectra shown in Fig. 6 were computed at the time when the magnetic energy density reached its maximum.
The spectral data were taken from Figs. 2 and 3 of the Ref. \citep{pol}. The dashed curves indicate the magnetic helicity-dominated spectrum Eq. (12)  for both cases: with fully helical (top) and non-helical (bottom) forcing. In the latter case, one can recognize the spontaneous parity violation caused by the chaotic/turbulent dynamics of the magnetized primordial plasma.\\

     Figure 7 shows a combined (Planck+ACT) CMB temperature power spectra obtained by the Planck mission and Atacama cosmology telescope (ACT). The CMB data shown in Fig. 7 were cleaned from the foreground by the Planck and ACT teams. The spectral data were taken from the corresponding sites\footnote{https://pla.esac.esa.int/pla and https://lambda.gsfc.nasa.gov/product/act/ }.  The ACT data were used for $kR > 1900$. The dashed curve indicates correspondence to the stretched exponential spectrum Eq. (12). 
 
\begin{figure} \vspace{-2.1cm}\centering 
\epsfig{width=.72\textwidth,file=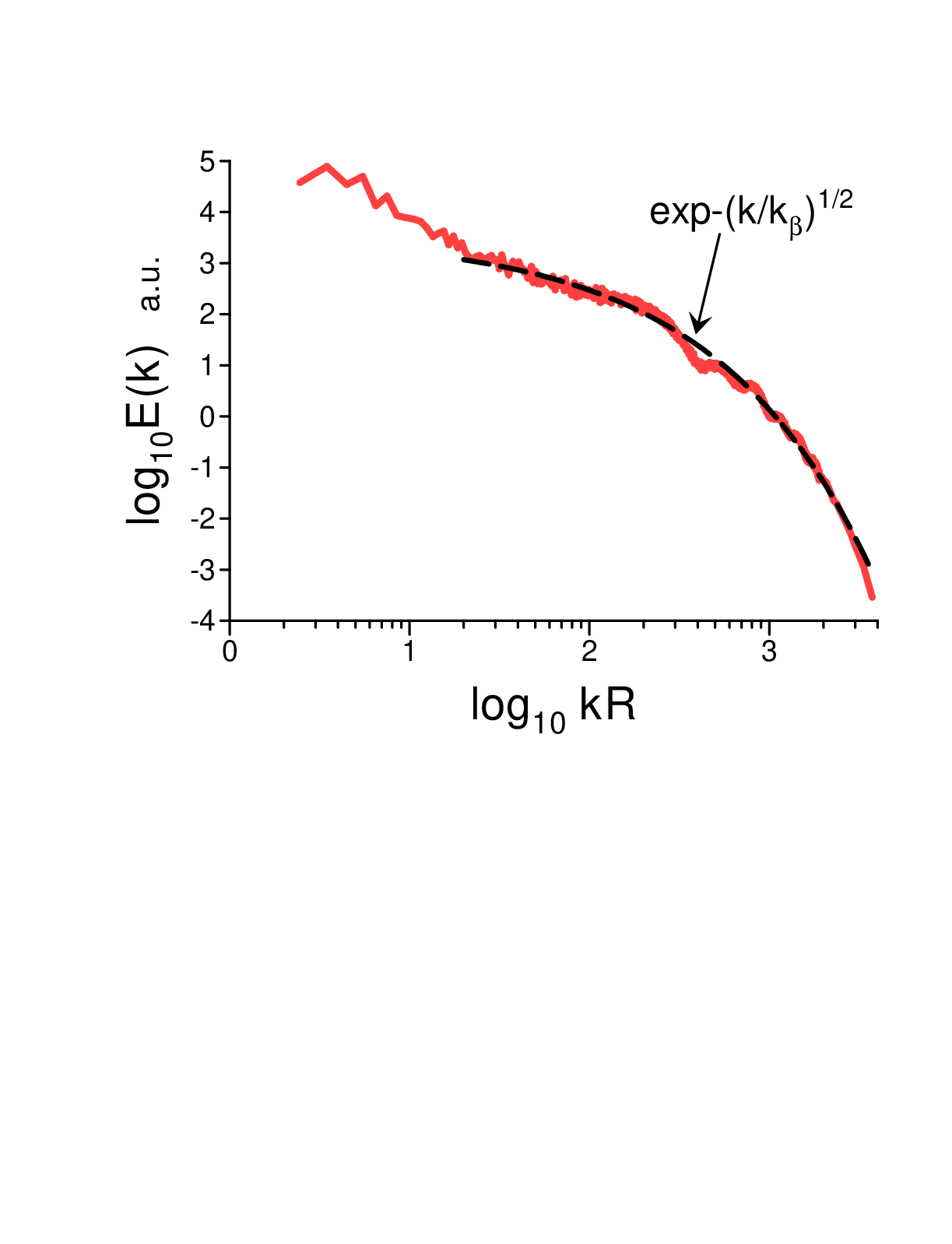} \vspace{-4.8cm}
\caption{Combined power spectrum of CMB temperature: Planck + ACT(for large values of $kR$).}
\end{figure}

     Figure 7 can be considered as an indication that the remnant spectrum from the lepto/baryogenesis epoch can indeed be seen in the CMB spectrum. The waviness of the spectrum can be related to the acoustic oscillations. The randomization with $\beta = 1/2$ provides evidence of the parity breaking and the magnetic helicity domination over the emission process of the CMB-to-be photons at this epoch. \\
     
      It is also interesting to look at the CMB temperature spectra obtained using ACTPol  98 and 150 GHz channels. Figure 8 shows these spectra on the log-linear scales. The spectral data were taken from the section `Multifrequency spectra' (`deep' files) at the site \citep{ACT} (see also \citealp{cho}). The dashed curve shows the best fit corresponding to Eq. (12) for the 150 GHz channel (for the 98 GHz channel, the best fit is about the same). The value of $k_{\beta}R \approx 16.5$ corresponds to that used in Fig. 7 for the combined Planck+ACT data. \\
     
     The CMB temperature with $\beta =1/2$ is considerably less randomized than the foreground with $\beta =1/4$, and $\beta =1/5$ (see Fig. 3 and below).

 \subsection{B-mode of CMB polarization fluctuations}    
 
\begin{figure} \vspace{-2.3cm}\centering 
\epsfig{width=.68\textwidth,file=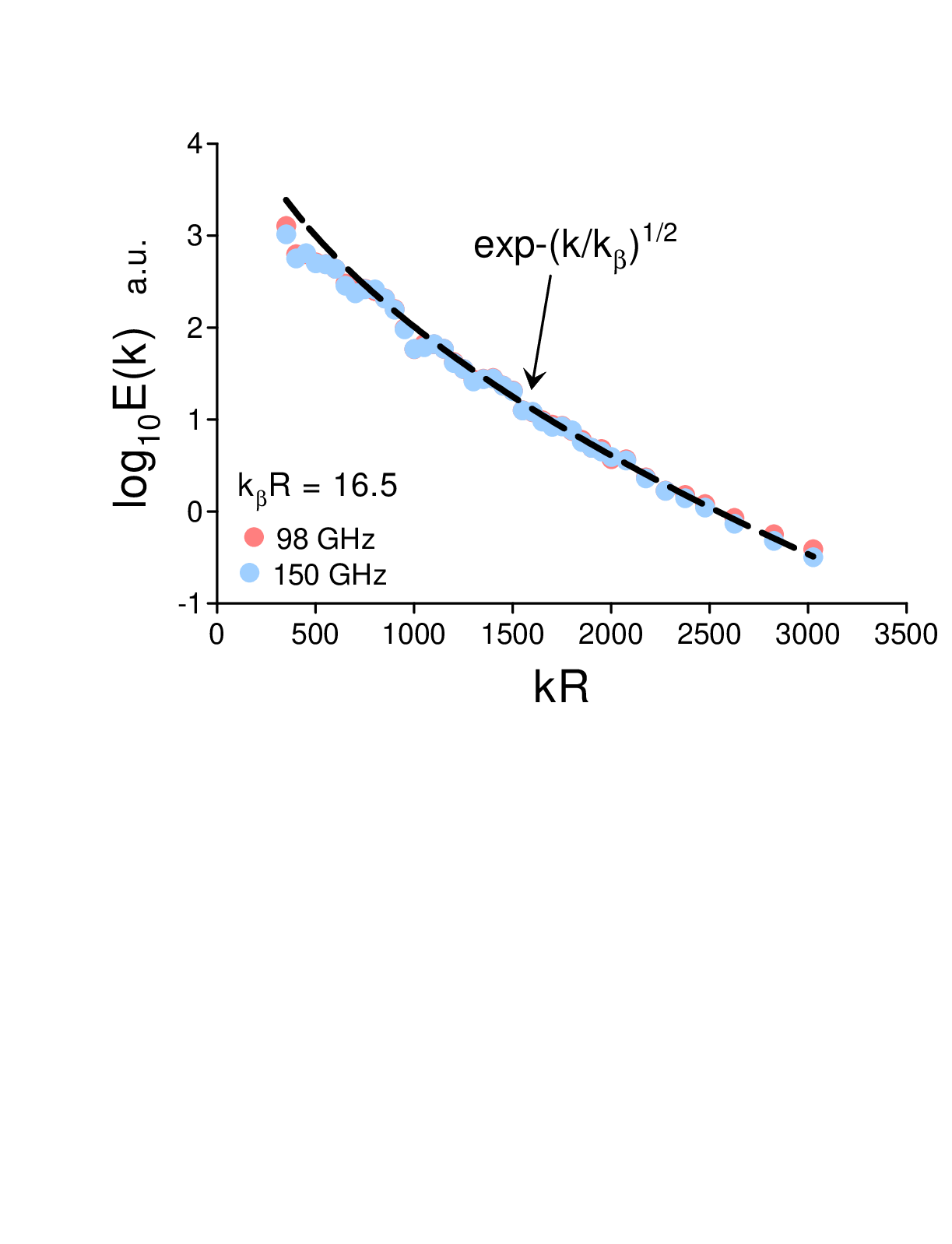} \vspace{-4.8cm}
\caption{Power spectra of CMB temperature fluctuations: the ACTPol measurements in 98 and 150 GHz channels.}
\end{figure}
\begin{figure} \vspace{-0.5cm}\centering  \hspace{-1cm} 
\epsfig{width=.69\textwidth,file=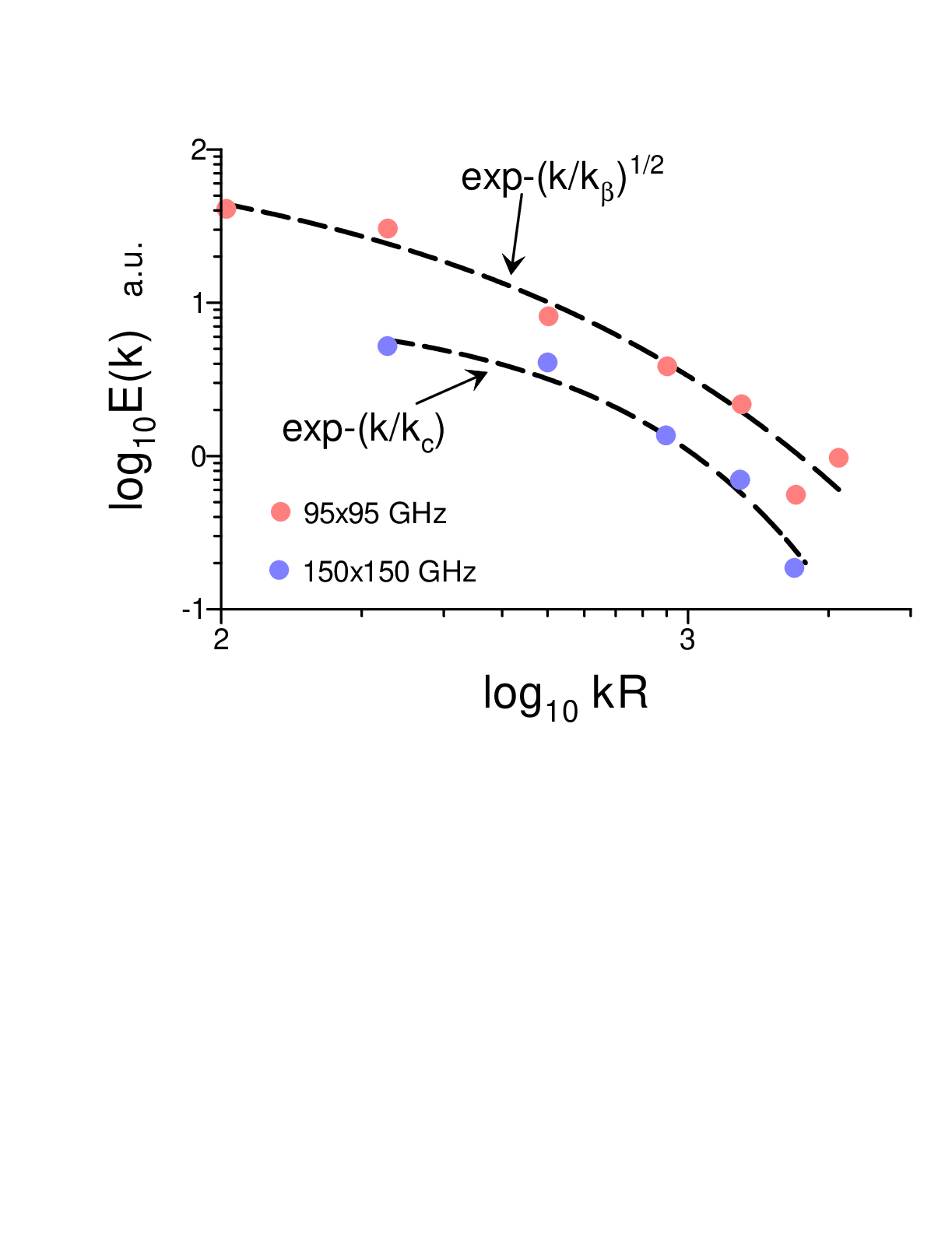} \vspace{-4.8cm}
\caption{Power spectra of B-mode of CMB polarization fluctuations: the South Pole Telescope (SPTpol) measurements.} 
\end{figure}
   In a Ref. \citep{zs} a spin-2 decomposition of the polarization tensor (vector) was suggested. In this technique, the polarization tensor (vector) is decomposed using two rotationally invariant scalar (E) and pseudo-scalar (B) quantities. The B-mode has magnetic-type parity.  
  
 The B-mode power spectrum observations are usually used to test theoretical/numerical models of the primordial magnetic fields. \\

 In a recent paper \citep{say}, results of measurements made with the South Pole Telescope (SPTpol) were reported.
This paper used 500 ${\rm deg}^2$ of the  SPTpol data set obtained for two frequency channels: 95 and 150 GHz. These are the most precise data for the CMB B-mode at large values of $l$ at present. \\

  Figure 9 shows B-mode bandpowers reported in the Ref. \citep{say} for the two frequency channels. The spectral data were taken from Table II of the Ref. \citep{say}. The dashed curves indicate the best fit corresponding to Eq. (12) (the distributed chaos with $\beta =1/2$) for the 95 GHz channel (top curve) and the best fit corresponding to Eq. (2) (the deterministic chaos) for the less powered channel 150 GHz (bottom curve), cf Fig. 5. \\

   One can see that the results obtained with the 95 GHz channel are in accordance with the CMB temperature measurements (Figs. 7 and 8). Since the results obtained with the less powered channel 150 GHz correspond to the deterministic chaos, it is interesting to compare them with the well-known $\Lambda$CDM multi-parametric model \citep{BPBB}. Figure 10 shows such a comparison in the log-linear scales. The best fit with Eq. (2) (the dashed straight line) was performed for the $\Lambda$CDM model. 
   
   It should be noted that the compact attractors in phase space corresponding to the deterministic chaos are usually determined by dissipation effects, whereas the distributed chaos (with $\beta = 1/2$) dominated by the adiabatic (quasi-ideal) helical invariant ${\rm I^{\pm}}$ should have a sufficient power to overcome the dissipation. 

\begin{figure} \vspace{-1.7cm}\centering  \hspace{-1cm} 
\epsfig{width=.65\textwidth,file=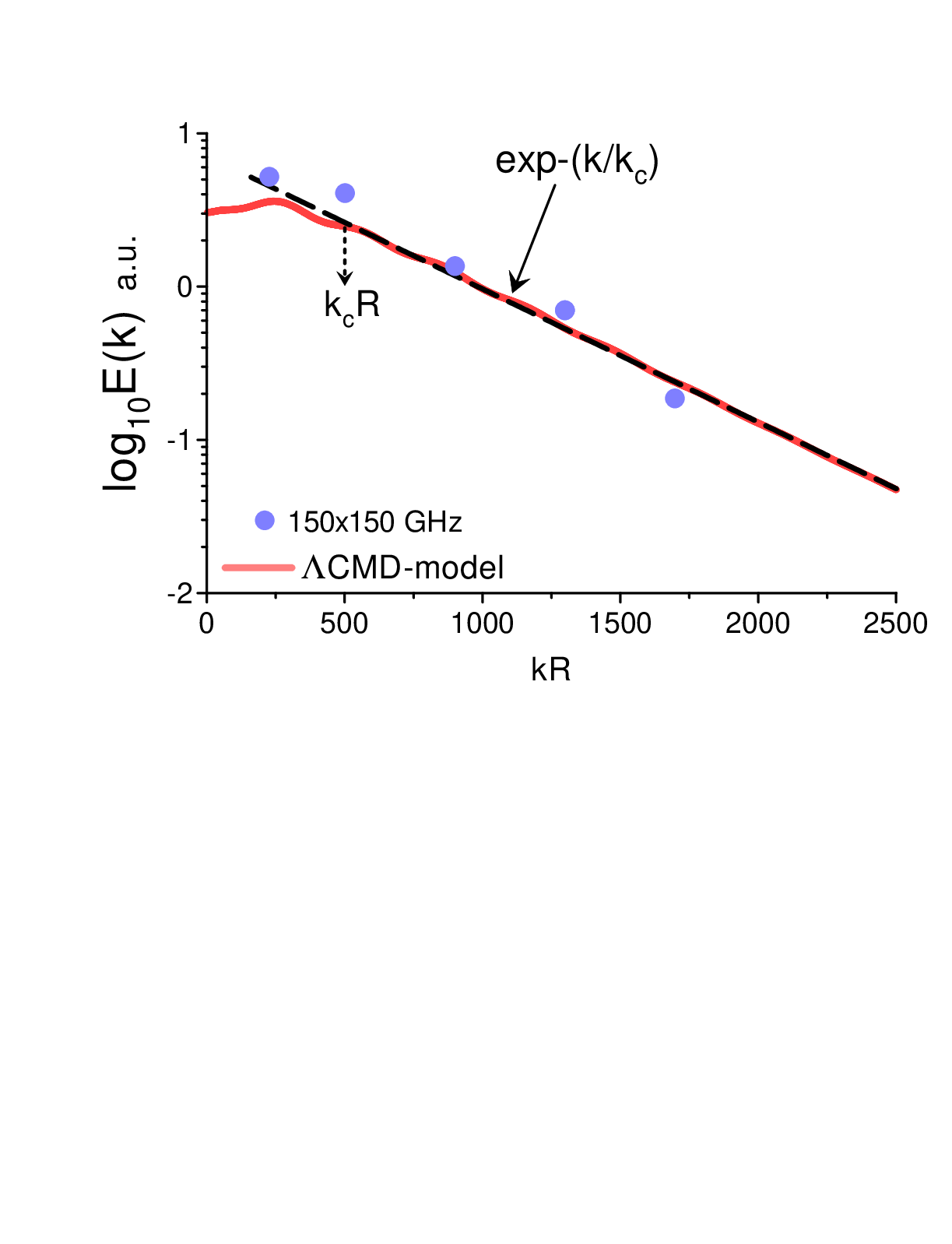} \vspace{-5cm}
\caption{Power spectra of B-mode of CMB polarization fluctuations: the 150 GHz channel of the SPTpol (circles) and the  $\Lambda$CDM multi-parametric model (red curve). The best fit with Eq. (2) (the dashed straight line) was performed for the $\Lambda$CDM model.} 
\end{figure}

\section{Magneto-inertial range of scales dominated by magnetic helicity}   

  For high Reynolds numbers the so-called inertial range of scales, dominated by the kinetic energy dissipation rate $\varepsilon$ only, is conventionally considered in hydrodynamic turbulence. The Kolmogorov phenomenology assumes that energy is transferred through this range with negligible dissipation to the sufficiently small scales where it is dissipated \citep{kol} (see also  \citealp{my} and references therein). In magnetohydrodynamics (and also at the kinetic scales) a magneto-inertial range of scales has been recently introduced in \citep{ber4}.  The two parameters: the magnetic helicity dissipation rate $\varepsilon_{h}$ and total energy dissipation rate $\varepsilon$ determine the magnetic field dynamics in this range. \\

   In the case of a considerable mean magnetic field the energy dissipation rate $\varepsilon$ can be replaced by the parameter $(\varepsilon \widetilde{B}_0)$ \citep{ir}. Here 
$\widetilde{B}_0 = B_0/\sqrt{\mu_0\rho}$ is normalized mean magnetic field. In the Alfv{\'e}n units the $\widetilde{B}_0$ has the same dimension as velocity. In Iroshnikov's phenomenology \citep{ir} the eddies, considered in the Kolmogorov phenomenology for hydrodynamics, are replaced by the Alfv{\'e}nic wave-packets which propagate in opposite directions along the mean magnetic field. The applicability of the Kolmogorov-like and Iroshnikov phenomenologies to magnetohydrodynamics was discussed for decades and some modifications were suggested. However, their main idea: using $\varepsilon$ or $(\varepsilon \widetilde{B}_0)$ as the dominant dimensional parameters in the inertial-like range of scales remained.  \\

    There is an analogy between the magneto-inertial range approach and the Corrsin-Obukhov inertial-convective range approach to the passive scalar where the two governing parameters: the passive scalar dissipation rate and energy dissipation rate, dominate the inertial-convective range (\citealt{my}) (see also \citealt{bs1}). \\
    
    According to this analogy, one can replace the estimate Eq. (10) with the estimate
\begin{equation}
 B_c \propto \varepsilon_{h_m}^{1/2} ~\varepsilon^{-1/6}~k_c^{1/6}  
\end{equation}
 for the magneto-inertial range without mean magnetic field, and with the estimate
\begin{equation}
 B_c \propto \varepsilon_{\hat{h}_m}^{1/2}~ (\varepsilon \widetilde{B}_0)^{-1/8}  k_c^{1/8}
\end{equation}

\begin{figure} \vspace{-2.5cm}\centering 
\epsfig{width=.7\textwidth,file=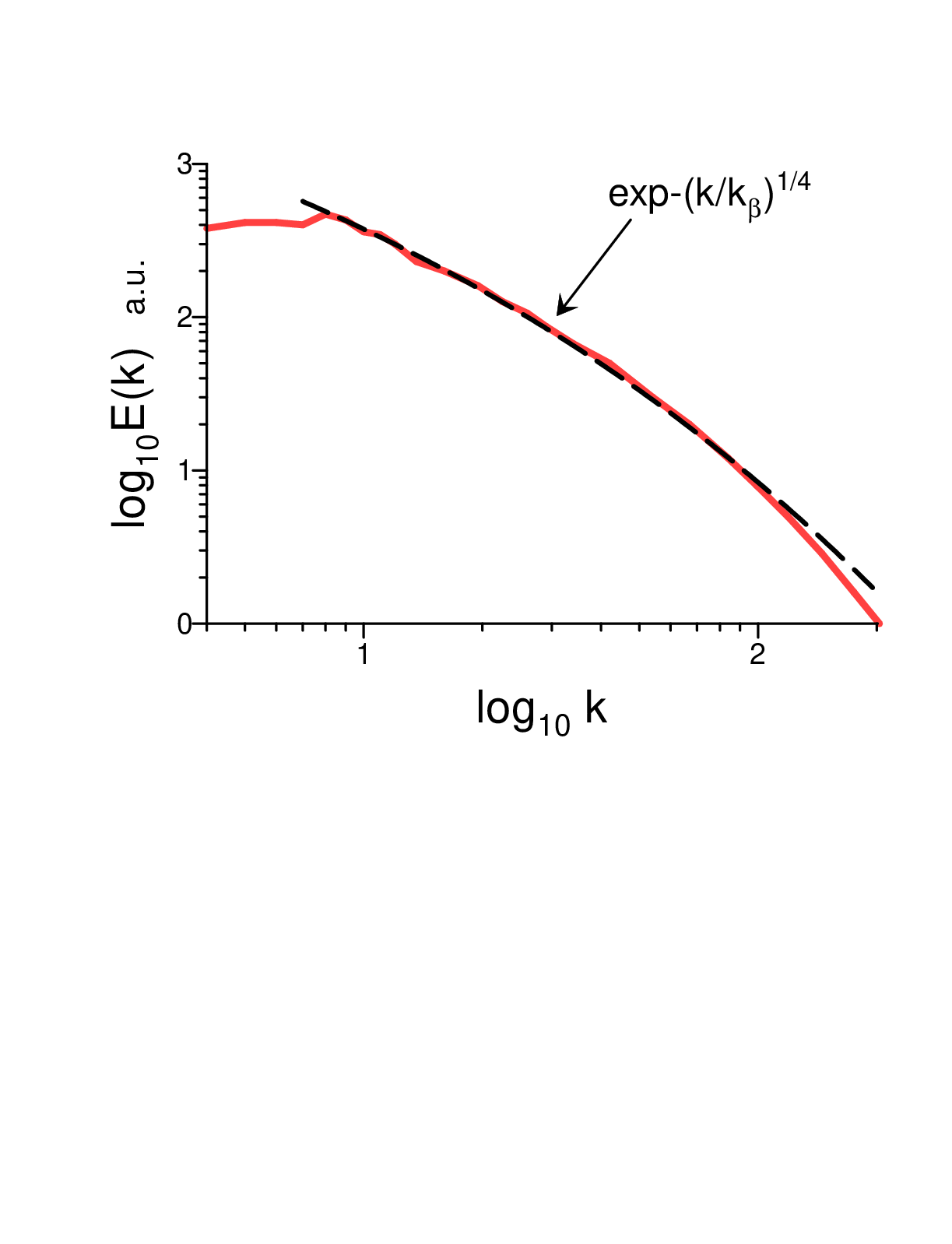} \vspace{-4.8cm}
\caption{Power spectrum of the chaotic/turbulent galactic magnetic field fluctuations (numerical simulations). } 
\end{figure}
 
     The specific estimates Eq. (10) and (15-17) can be generalized
\begin{equation}
 B_c \propto k_c^{\alpha}   
\end{equation}
  
  In the asymptotic of large $k_c$ the stretched exponential form of the distributed chaos spectra 
\begin{equation}
 \int_0^{\infty}  P(k_c) \exp -(k/k_c)dk_c \propto \exp-(k/k_{\beta})^{\beta} 
\end{equation}  
results in the probability distribution \citep{jon}
\begin{equation}
P(k_c) \propto k_c^{-1 + \beta/[2(1-\beta)]}~\exp(-\gamma k_c^{\beta/(1-\beta)}) 
\end{equation}   
  
  In the case of the half-normally distributed $B_c$ a relationship between $\alpha$ and $\beta$ can be obtained from the Eqs. (18) and (20)
\begin{equation}
\beta = \frac{2\alpha}{1+2\alpha}  
\end{equation}

\begin{figure} \vspace{-2.5cm}\centering 
\epsfig{width=.7\textwidth,file=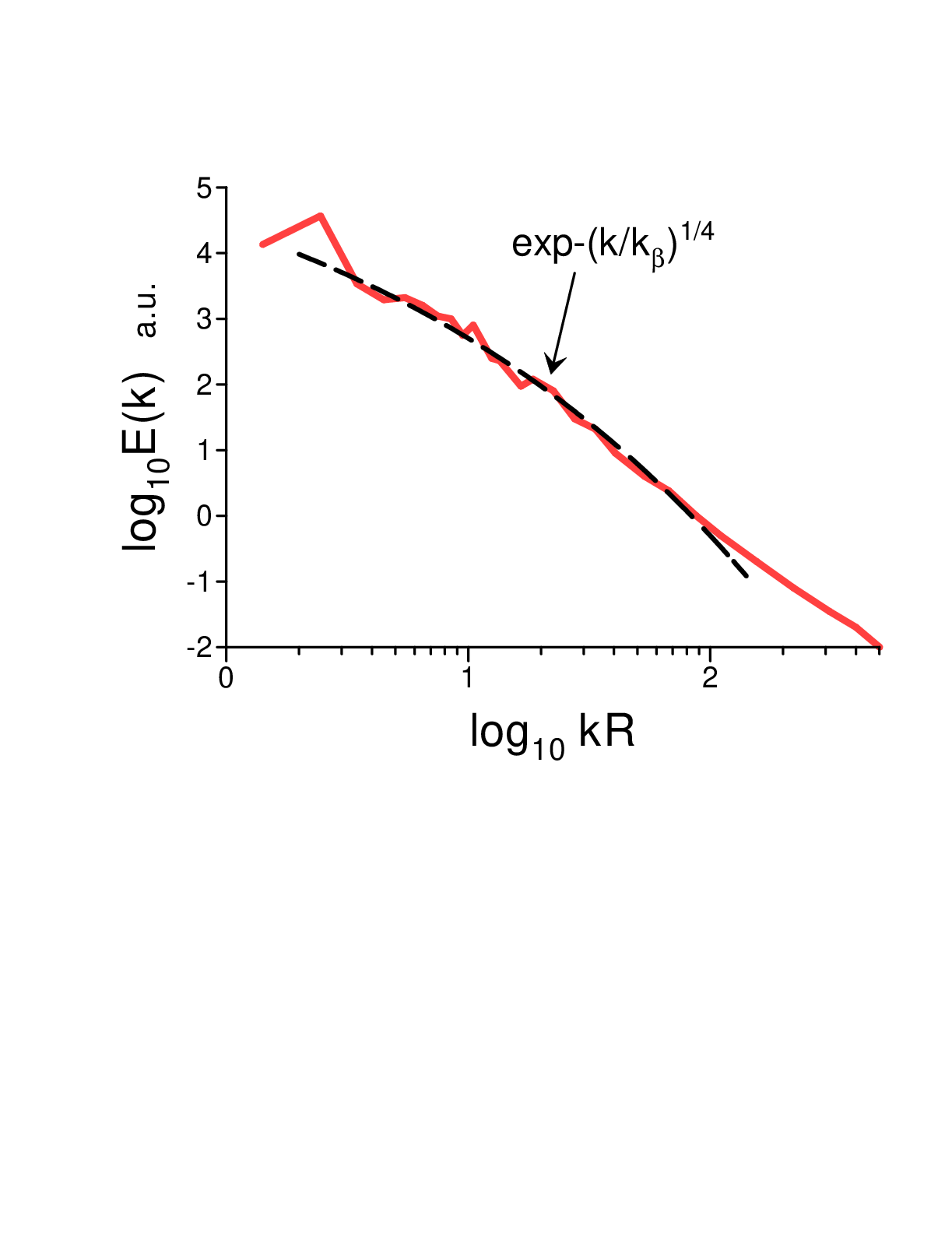} \vspace{-4.6cm}
\caption{Power spectrum of the average magnetic field $B_{LoS}$ all-sky map (obtained from the Faraday rotation all-sky map).} 
\end{figure}

 For $\alpha =1/6$ Eq. (16), we obtain from Eq. (21)
\begin{equation}
 E(k) \propto \exp-(k/k_{\beta})^{1/4} , 
\end{equation}
 and for $\alpha =1/8$ Eq. (17), we obtain from Eq. (21)
\begin{equation}
 E(k) \propto \exp-(k/k_{\beta})^{1/5} 
\end{equation} 

   A numerical simulation of a MHD dynamo in a Milky Way-like galaxy was performed in a recent article \citep{nto} using equations:
\begin{eqnarray}
\frac{\partial\rho}{\partial t} +\nabla(\rho{\bf u})& =& 0 \\
\frac{\partial\rho{\bf u}}{\partial t} + \nabla\cdot\left(\rho\bf{u}\bf{u}-\bf{B}\bf{B}\right) + \nabla P_{tot} &=& -\rho\nabla\phi  \\
\frac{\partial E_{tot}}{\partial t} +\nabla\left[~(E_{tot}+P_{tot})\bf{u} -(\bf{u}\cdot\bf{B})\cdot\bf{B}\right] &=& -{\bf u}\cdot\nabla\phi -\rho\Lambda + \Gamma \\
\frac{\partial\bf{B}}{\partial t} -\nabla\times(\bf{u}\times\bf{B}) &=& 0  \\
\nabla\cdot\bf{B} &=& 0 
\end{eqnarray}
  The functions of density $\rho$ and temperature $T$: $\Gamma=\Gamma(\rho,T)$ and  $\Lambda=\Lambda(\rho,T)$, represent the heating and cooling rates of the intragalactic plasma, $\phi$ is the gravitational potential, the total pressure was taken as
\begin{equation}
    P_{tot} = p + \frac{\bf{B}\cdot\bf{B}}{2}
\end{equation}
and the total energy was taken as
\begin{equation}
    E_{tot} = E_{int} + \rho\frac{\bf{u}\cdot\bf{u}}{2} +  + \frac{\bf{B}\cdot\bf{B}}{2}
\end{equation}
with $E_{int}$ being the internal energy of the fluid. The equation of state was taken as $P=(\gamma-1)E_{int}$. \\ 

\begin{figure} \vspace{-2.4cm}\centering 
\epsfig{width=.7\textwidth,file=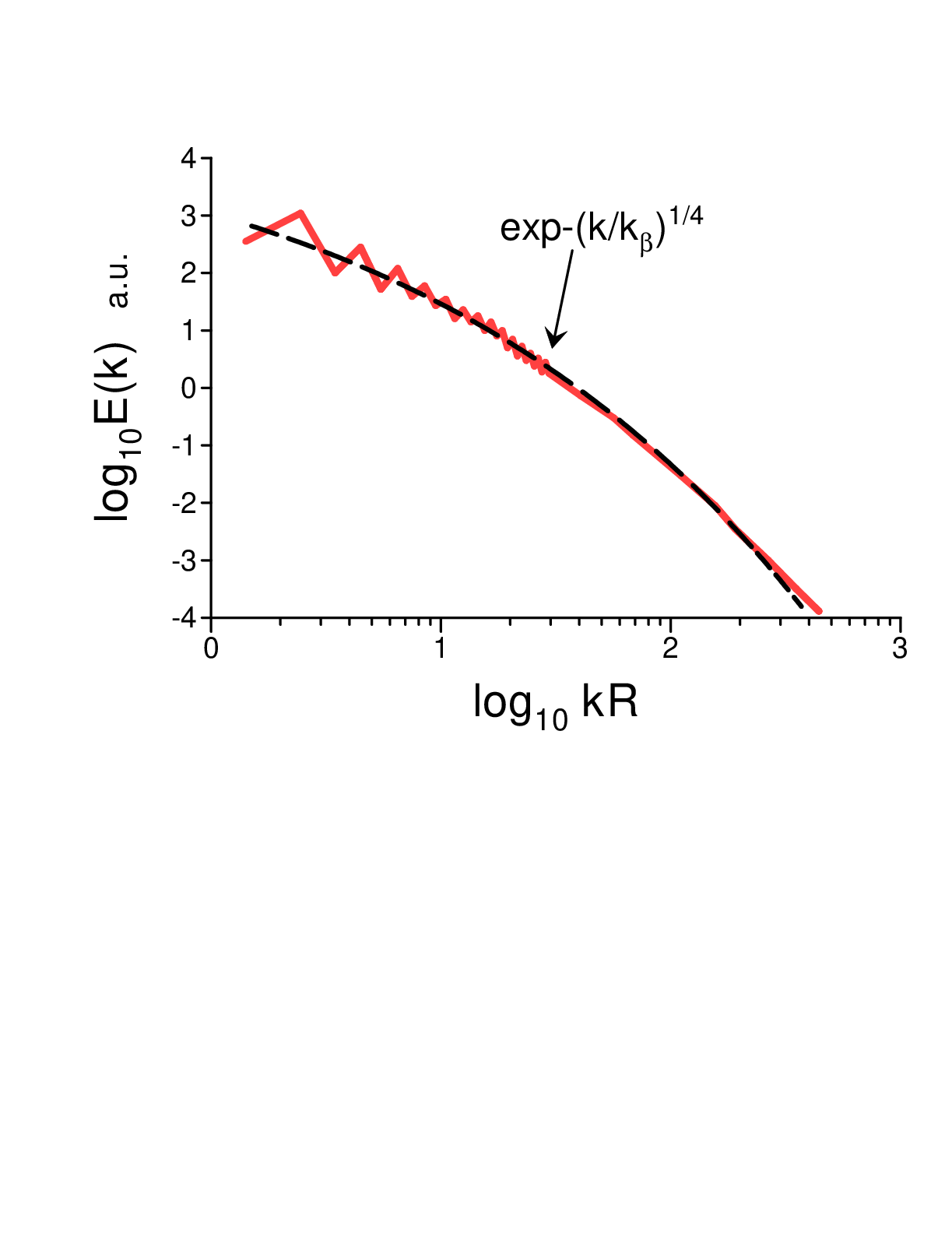} \vspace{-4.8cm}
\caption{Power spectrum of electron dispersion measure all-sky map inferred from the Faraday rotation map.} 
\end{figure}

  The model also takes into account the stars, their feedback, and the dark matter halo.  The configuration of the intragalactic medium simulates a Milky Way-like galaxy. As a seed initial condition a weak toroidal magnetic field was taken. The chaotic motion of the intragalactic plasma was generated (simulated) by the galactic differential rotation and supernova explosions. At some stage of the simulation, the generated magnetic field was smoothed to separate the mean field and chaotic residual fluctuations.  
  
  Figure 11 shows the magnetic energy spectrum of chaotic magnetic field fluctuations computed at this stage. The spectral data were taken from Fig. 5 of the Ref. \citep{nto}. The dashed curve indicates the best fit by the stretched exponential Eq. (22) (the magneto-inertial range of scales).\\

\section{Observations in the Galactic magnetized plasma}      

 The main difficulty of measurements of the interstellar magnetic field is that (unlike the interplanetary magnetic field) it cannot be measured directly. In the accessible observables, the magnetic field is usually entangled and mixed with other variables characterizing the plasma.
 
  The Faraday effect, for instance, is a rotation of the polarization position angle propagating through the magnetized plasma. It entangles the electron density $n_e$ with the line-of-sight component of the magnetic field $B_{LoS}$ into an observable characteristic - Faraday depth:
\begin{equation}
\phi  = \frac{e^3}{2\pi m_e^2 c^4}\int_{\mathrm{LoS}} dl\, n_{\mathrm{e}} B_{LoS} 
\end{equation}

   In the paper (\citealt{hut}) a disentangling of an all-sky map for average intragalactic $B_{LoS}$ from the Faraday effect map was reported. Some complementary tracers of the $n_e$ (the free-free map of the Planck survey \citealp{Planck}, extra-Galactic Faraday data \citealp{eak}, a $H-\alpha$ map \citealp{fin}, pulsar data \citealp{man}) were also used for the purpose.  An all-sky map of the electron dispersion measure (the integrated electron density) was also constructed. \\

\begin{figure} \vspace{-1.5cm}\centering 
\epsfig{width=.65\textwidth,file=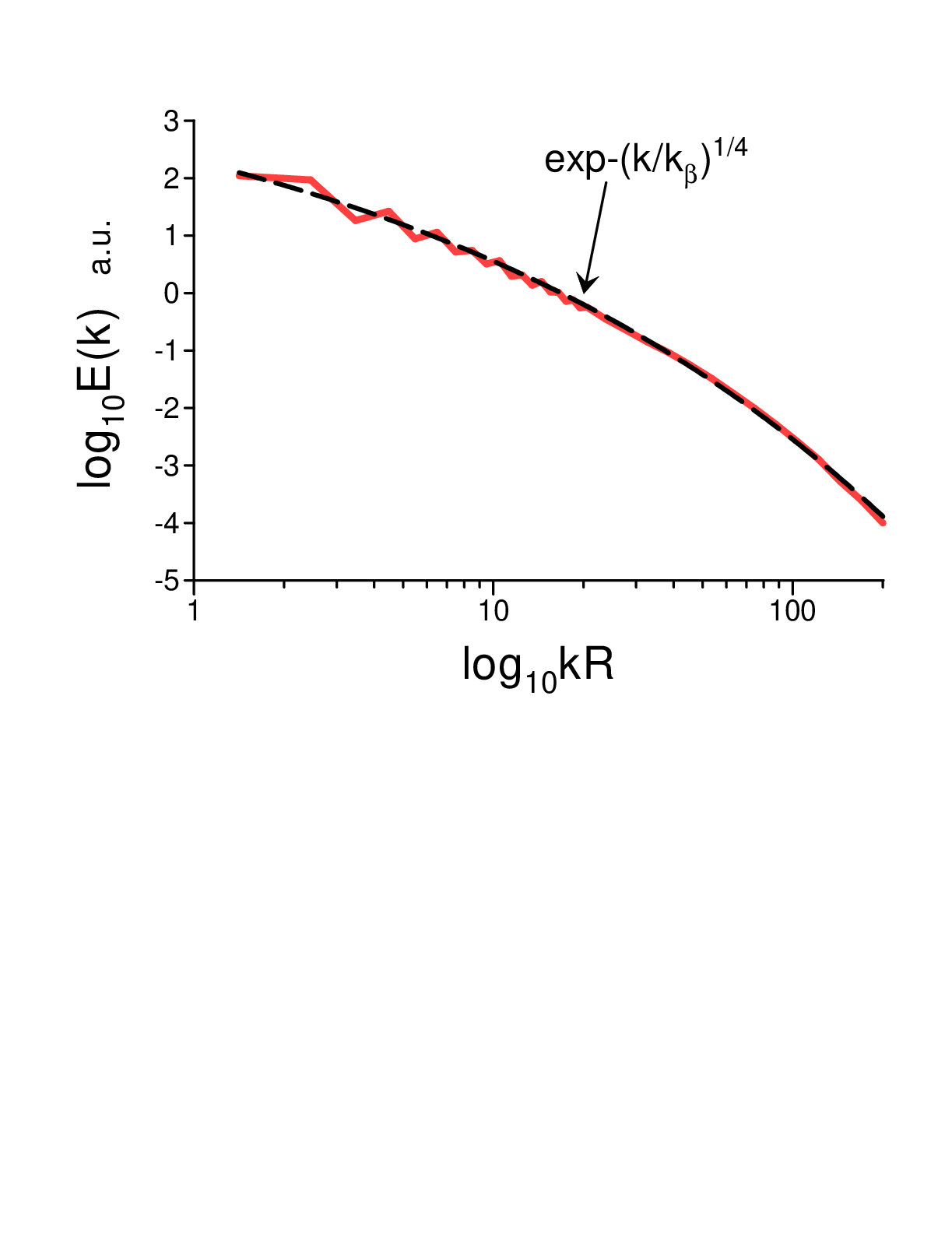} \vspace{-4.8cm}
\caption{Power spectrum of the all-sky synchrotron emission (the Haslam 0.408 GHz survey).} 
\end{figure}

     Figure 12 shows the power spectrum of $B_{LoS}$. The spectral data were taken from Figs. 9a of the paper (\citealt{hut}). The dashed curve in Fig. 12 indicates the best fit by the stretched exponential spectral law Eq. (22). \\
     
    An all-sky map of the integrated electron density (electron dispersion measure) was also obtained by this method. Figure 13 shows the power spectrum of the electron dispersion measure. The spectral data were taken from Figs. 9b of the paper (\citealt{hut}). The dashed curve in Fig. 13 indicates the best fit by the same stretched exponential spectral law Eq. (22). Apparently the magnetic field imposes its degree of randomization (the $\beta =1/4$) on the electron dispersion measure.

\section{Synchrotron emission}

  The astroparticles (relativistic electrons and positrons) produce the synchrotron emission moving through the magnetic field. Therefore, the diffuse (polarized) synchrotron emission is an effective tracer of the magnetic field in the magnetized non-thermal plasma. This emission is also one of the main components of the polarized foreground for the cosmic microwave background radiation. Therefore,  studying its spectral properties is important for understanding the magnetized intragalactic plasma's physical processes and obtaining clean CMB maps.\\
  
\begin{figure} \vspace{-2.4cm}\centering 
\epsfig{width=.7\textwidth,file=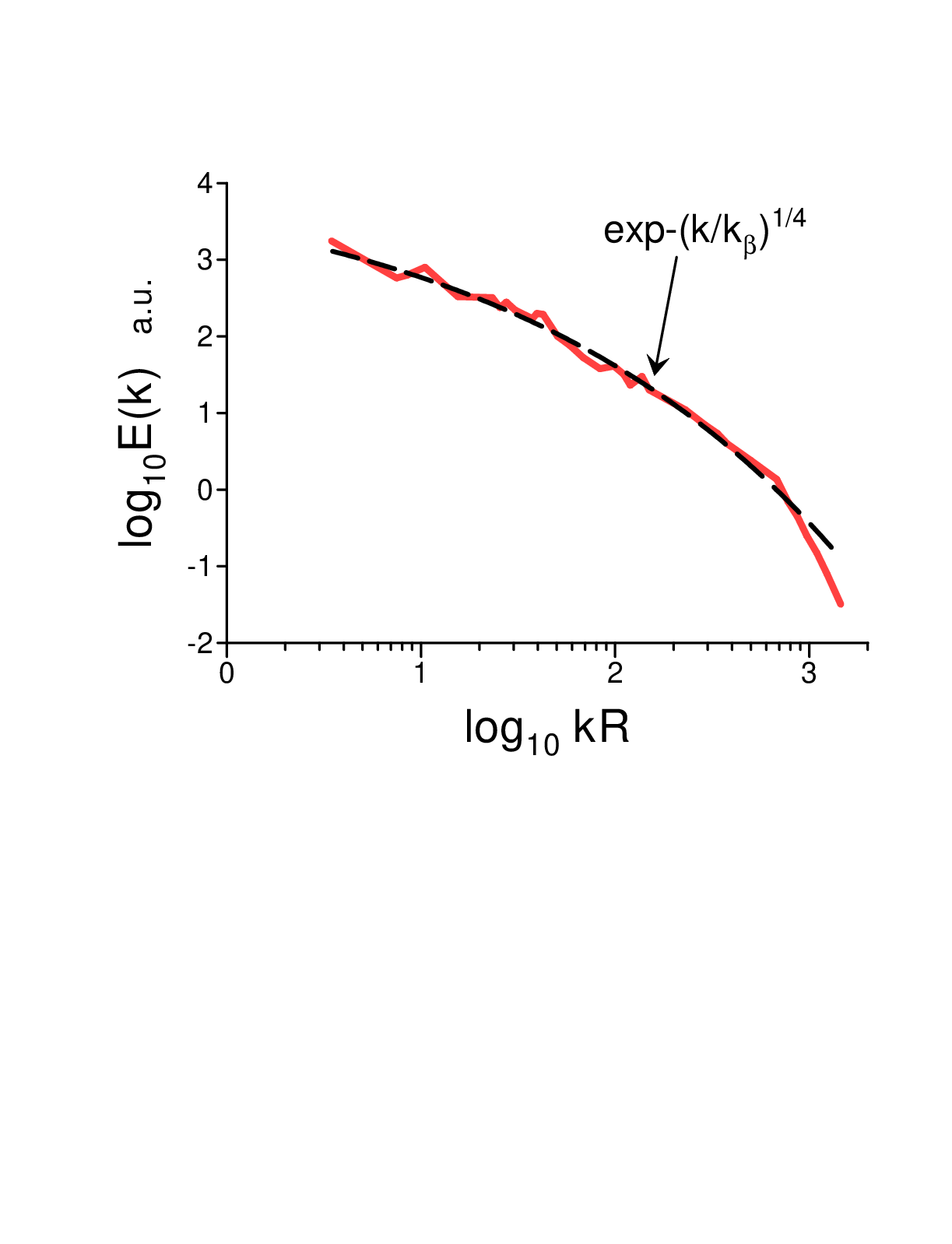} \vspace{-4.6cm}
\caption{Power spectrum of the B-mode of synchrotron polarised emission of the Southern Galactic plane (the Parkes 2.4 GHz survey).} 
\end{figure}
  
     In a paper \citep{mer} a reanalysis of the famous so-called  `Haslam' 0.408 GHz all-sky survey of the synchrotron emission was made, with a subtraction of the point sources and an averaging over different lines-of-sight.  Figure 14 shows the power spectrum obtained for this reconsidered Haslam survey. The spectral data were taken from Fig. 3 of the Ref. \citep{mer}. 
      
   The dashed curve in Fig. 14 indicates the best fit by the stretched exponential spectral law Eq. (22). Apparently the magnetic field imposes its degree of randomization in this case as well.\\
  
   In a paper \citep{gia} results obtained using this technique for the Parkes 2.4 GHz survey of synchrotron polarised emission for the Southern Galactic plane were reported. Figure 15 shows the power spectrum for the B-mode. The spectral data were taken from Fig. 3 of the Ref. (\citealt{gia}). The dashed curve in Fig. 15 indicates the best fit by the stretched exponential spectral law Eq. (22).

\section{Dust emission}  

\begin{figure} \vspace{-2.5cm}\centering 
\epsfig{width=.65\textwidth,file=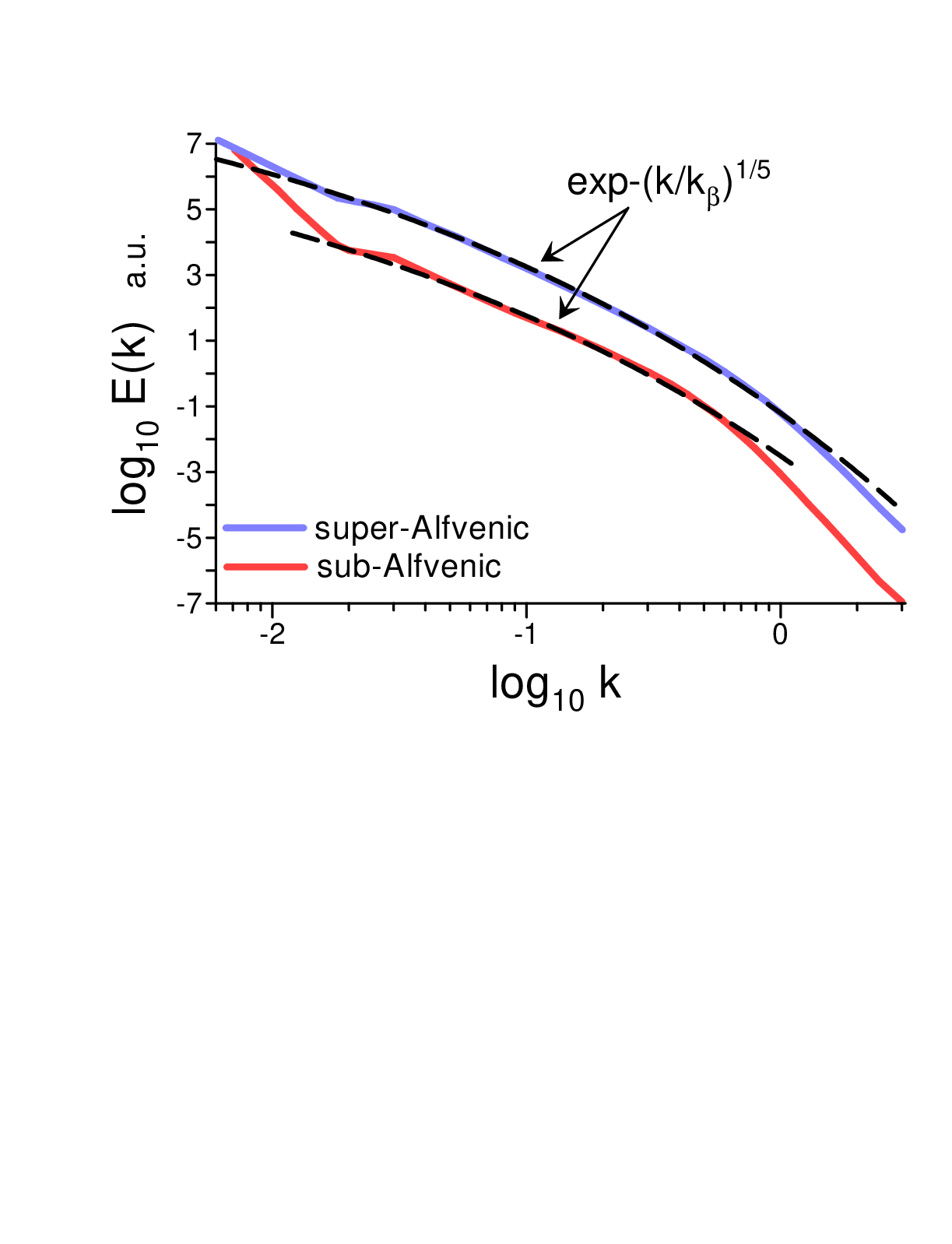} \vspace{-4.7cm}
\caption{Power spectra of sub-Alfv\'{e}nic (bottom) and super-Alfv\'{e}nic (top) magnetic field. The sonic Mach number $M  \approx 6$ for both cases (numerical simulation).} 
\end{figure}
\begin{figure} \vspace{-0.3cm}\centering 
\epsfig{width=.6\textwidth,file=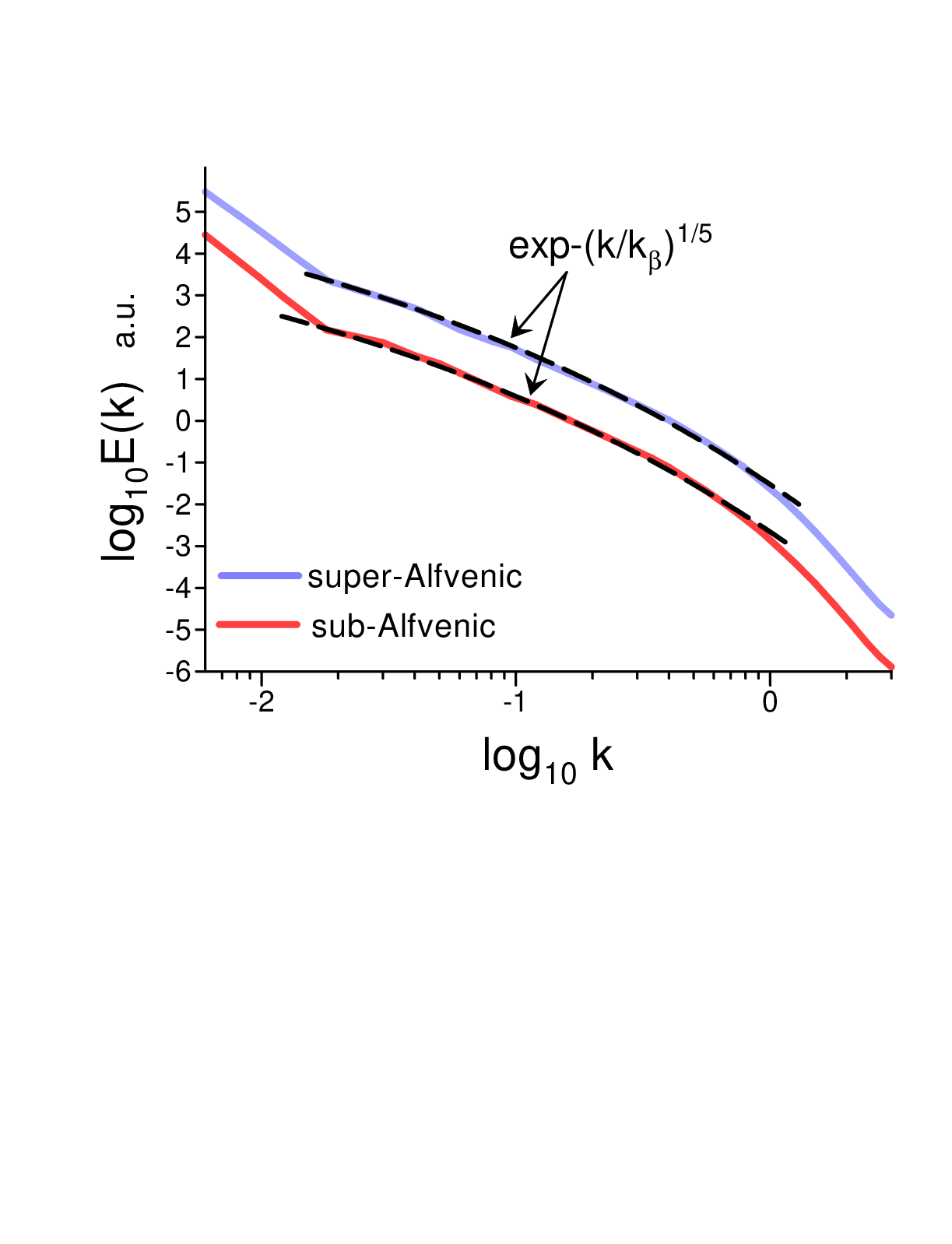} \vspace{-3.6cm}
\caption{As in Fig. 16. but for the T-mode of the thermal polarized dust emission} 
\end{figure}
\begin{figure} \vspace{-0.3cm}\centering 
\epsfig{width=.61\textwidth,file=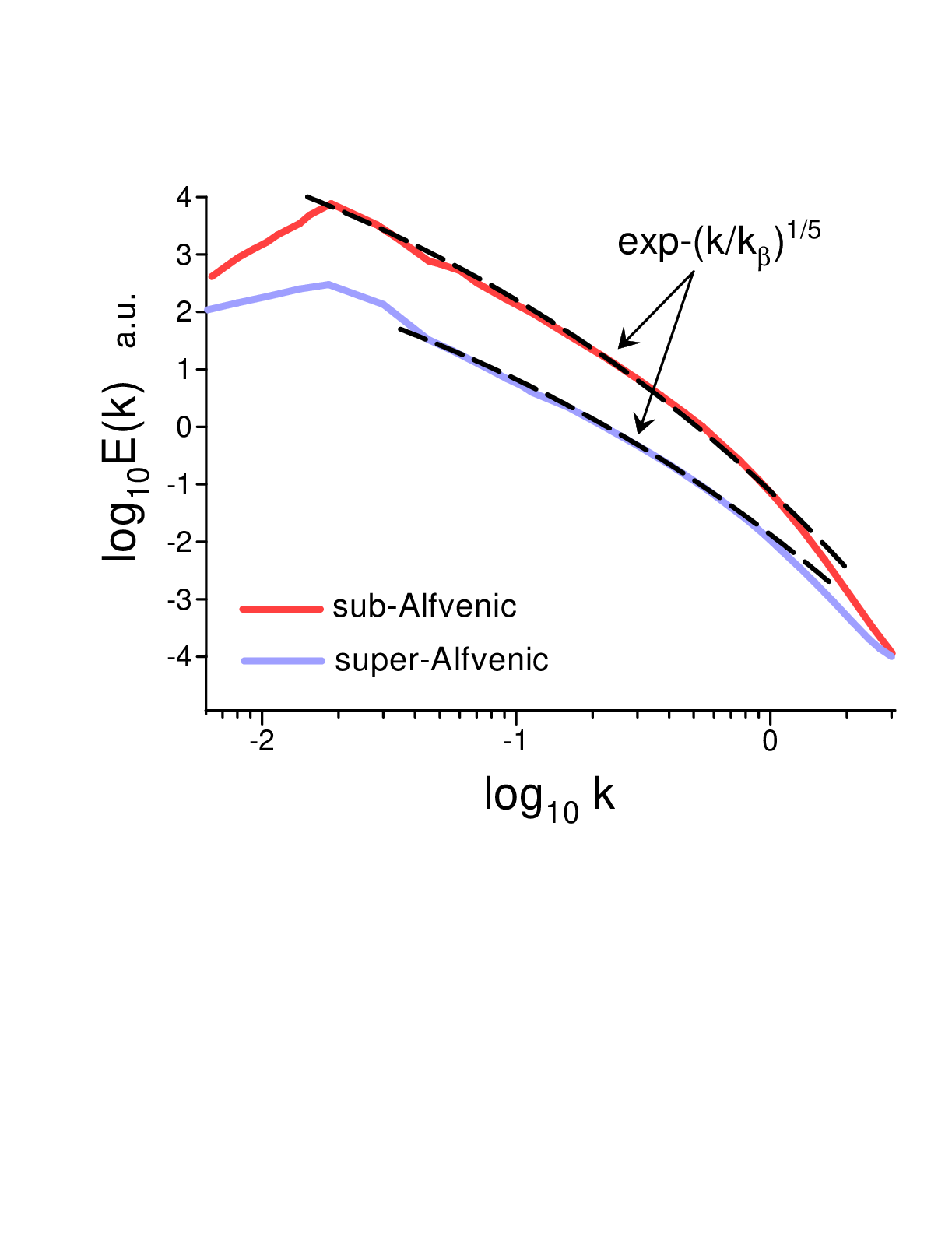} \vspace{-3.3cm}
\caption{As in Fig 16 but for the B-mode of the thermal polarized dust emission} 
\end{figure}
 
  Polarized dust emission is not only one of the most reach sources of information about the Galactic magnetic field (see, for instance, a recent paper by \citealp{hl}, and references therein) but also the main obstacle for observing (detecting) the gravitational waves in primordial CMB B-modes, especially for the high-frequency Planck's channels \citep{kog}.\\

  Let us begin with a recent numerical simulation. Results of this simulation were reported in Ref. \citep{sch}.  The ideal compressible MHD simulation was performed in a 3D periodic spatial box with an approximately isothermal equation of state. A stochastic (Gaussian) non-helical large-scale forcing with a constant energy injection rate drove the gas. The polarization of the thermal dust emission comes from the elongated grains of dust grains spinning around the local magnetic
field (their long axes were perpendicular to the field). The dust-to-gas ratio was uniform and constant, the grains were aligned with the magnetic field direction, the
dust cloud was optically thin, and the dust temperature coincided with the constant gas temperature. The projections perpendicular to the mean magnetic field were the main subject of the consideration. The power spectra were computed by averaging over an annulus in Fourier space.\\

  Figure 16 shows the power spectra of sub-Alfv\'{e}nic (bottom) and super-Alfv\'{e}nic (top) magnetic field. The sonic Mach number $M  \approx 6$ for both cases. The spectral data were taken from Fig. 4 of the Ref. \citep{sch}.
  
\begin{figure} \vspace{-2.5cm}\centering 
\epsfig{width=.65\textwidth,file=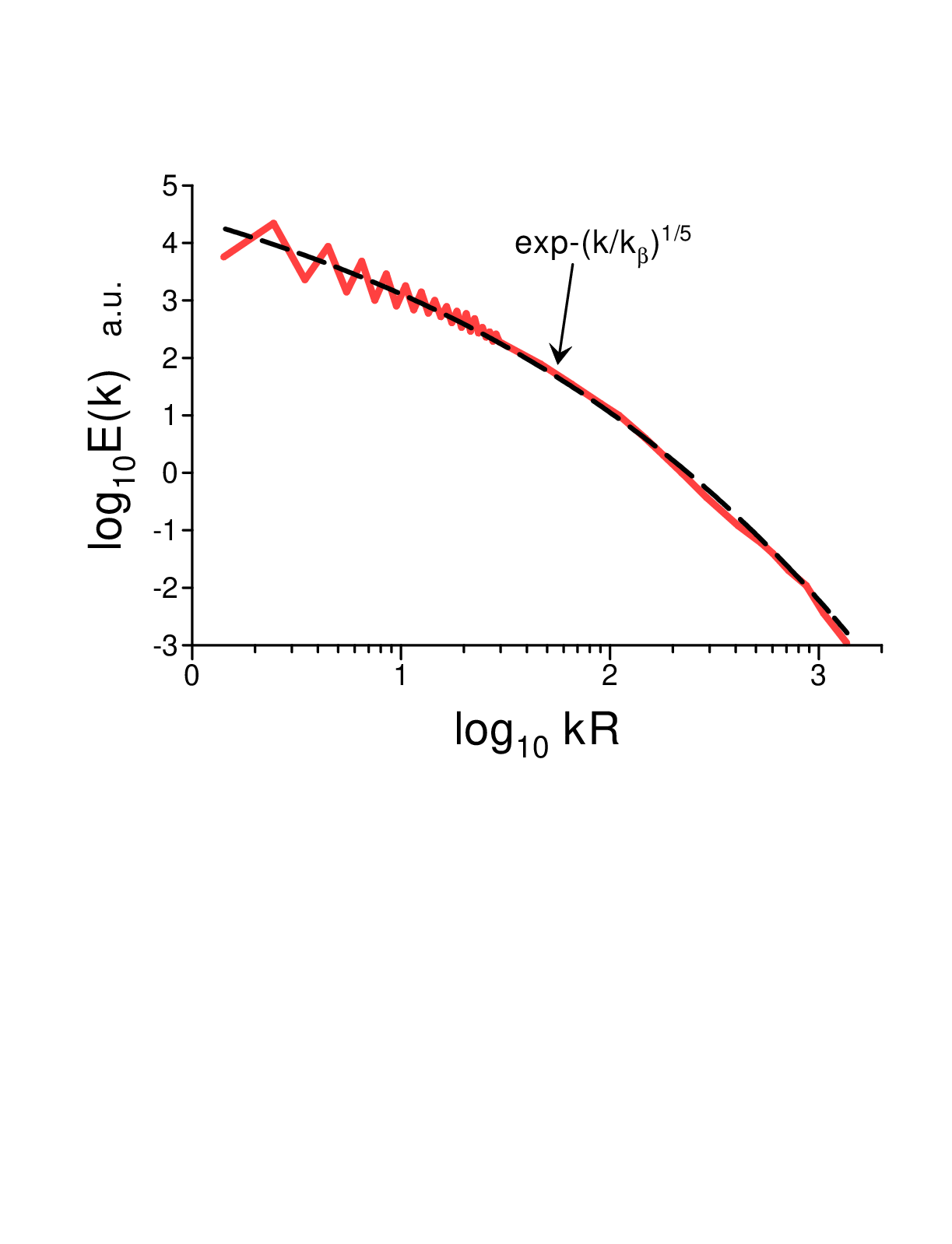} \vspace{-4.2cm}
\caption{Power spectrum for the T-mode of the polarized dust emission computed for GNILC cleaned full-sky maps obtained using the Planck data at 353 GHz. } 
\end{figure}
  
  Figures 17 and 18 show the projected T-mode (temperature ) and B-mode (polarization) power spectra of the polarized dust emission for the same conditions. The spectral data were taken from Fig. 6 of the Ref. \citep{sch}. 
  
  The dashed curves in Figs. 16,17, and 18 indicate the best fit by the stretched exponential spectral law Eq. (23), that means the importance of the mean magnetic field in this simulation. The same level of randomization of the magnetic field and the polarized emission confirms the imposition of the magnetic field level of randomization on the polarized emission (cf previous Section).    \\
  
   Figures 19 and 20 show the T-mode (temperature ) and B-mode (polarization) power spectra of the polarized dust emission computed using the GNILC-cleaned full-sky maps obtained using the Planck data at 353 GHz. The spectral data were taken from Fig. 11 of a paper \citep{mkd}. The dashed curves in Figs. 19 and 20 indicate the best fit by the stretched exponential spectral law Eq. (23) (cf Figs. 17 and 18).\\

\begin{figure} \vspace{-2.5cm}\centering 
\epsfig{width=.67\textwidth,file=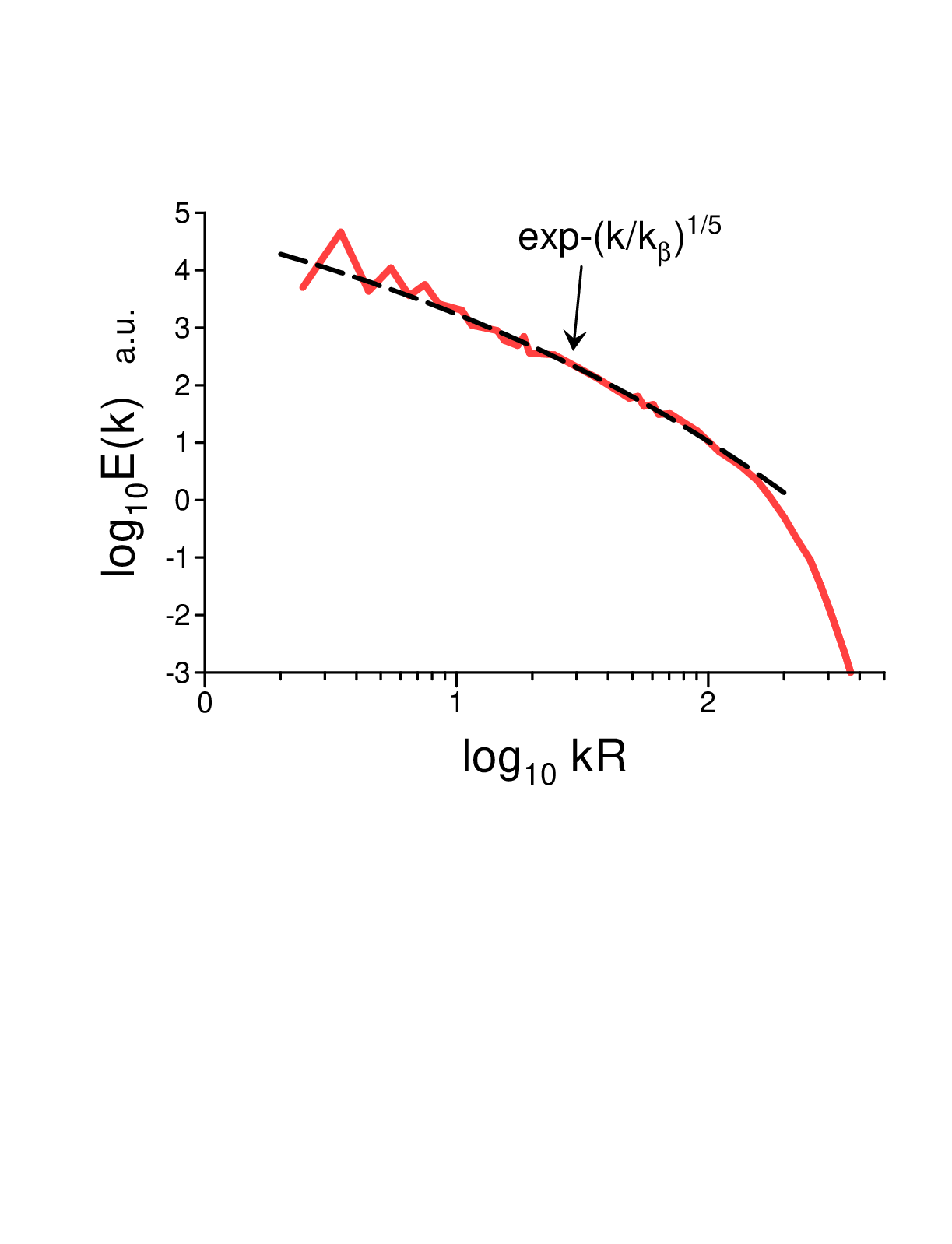} \vspace{-4cm}
\caption{Power spectrum for the B-mode of the polarized dust emission computed for GNILC cleaned full-sky maps obtained using the Planck data at 353 GHz. } 
\end{figure}
\begin{figure} \vspace{-0.5cm}\centering 
\epsfig{width=.67\textwidth,file=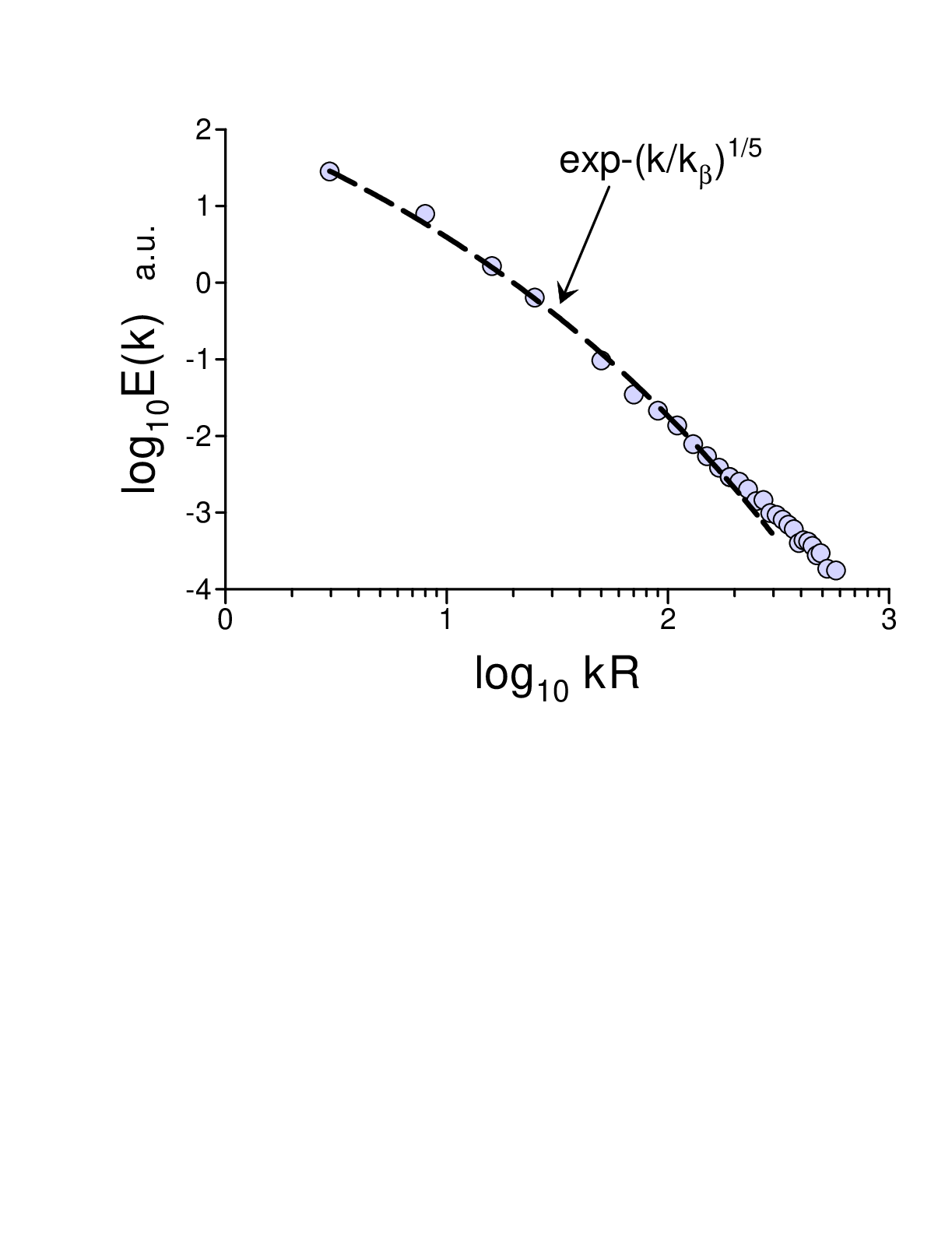} \vspace{-4.8cm}
\caption{Power spectrum of the B-mode of the polarized dust emission computed for a 71\% sky map obtained using the Planck data at 353 GHz. } 
\end{figure}
   In a recent paper \citep{akr} the third  Planck public release's maps were used to characterize the polarized dust emission at high Galactic latitudes. Using these maps the angular power spectra of the B-mode were computed at 353 GHz channel for six sky regions. The largest region covers 71 \% of the sky (naturally, the best measurement results were obtained in this region). Figure 21 shows the power spectrum of the B-mode for this region. The spectral data were taken from Table C.1 of the Ref. \citep{akr}. The dashed curve in Fig. 21 indicates the best fit by the stretched exponential spectral law Eq. (23).\\
   
 \section{Conclusions and Discussion}
 
  The galactic foreground emission becomes more sensitive to the mean magnetic field with frequency. One can assume that the two levels of randomization observed in Fig. 3: lower ($\beta=1/4$) for the smaller frequencies and higher ($\beta=1/5$) for larger frequencies, can be related to the difference in the spatial location of the Galaxy regions providing the main contribution to the foreground emission at different frequencies. Namely, in the regions providing the main contribution to the K and Ka channels the mean magnetic field does not play an important role (can be neglected) and therefore the spectra should be described by Eq. (22) whereas in the regions providing the main contribution to the Q,V, and W channels the mean magnetic field plays an important role and the spectra should be described by Eq. (23). 
 
   Analogously, one can conclude that in the regions providing the main contribution to the all-sky map for the Faraday rotation measure the mean magnetic field also does not play an important role. Therefore, the spectra corresponding to the magnetic field $B_{LoS}$ Fig. 12 and integrated electron density (electron dispersion measure) Fig. 13, obtained by disentangling the all-sky Faraday rotation map, are well fitted by Eq. (22). \\
   
   The difference in the levels of randomization could also be partially related to the fact that for the smaller and middle frequencies the synchrotron and free-free emissions dominate the Galactic foreground whereas for the large frequencies the dust emission takes this role (see Fig. 5  of Ref. \citealp{kog}). The problem is that the observed angular power spectra of the synchrotron and free-free emission for the middle and high frequencies are rare in the literature as well as the spectra of the dust emission for small frequencies. Maybe the above consideration can be encouraging to solve this problem. \\

  The results of the numerical simulations and the astrophysical observations indicate that the magnetic helicity-dominated magnetic field imposes its level of randomization on the synchrotron and dust emission.\\
  
  The background radiation is considerably less randomized  ($\beta =1/2$) than the Galactic foreground radiation ($\beta =1/4$, and $\beta =1/5$). This difference can be used for effective cleaning of the raw CMB data from the foreground.\\
   
   Despite the vast differences in the values of physical parameters and spatio-temporal scales between the numerical simulations and the astrophysical observations, there is a quantitative agreement between the results of the astrophysical observations and the numerical simulations in the framework of the distributed chaos approach.

\end{document}